\documentclass[prl,twocolumn, aps,amssymb,nofootinbib,longbibliography,showpacs,superscriptaddress]{revtex4-2}

\usepackage{graphicx}
\usepackage{dcolumn}
\usepackage{siunitx}
\usepackage{hyperref}
\usepackage{mathtools}
\usepackage{svg}
\usepackage{float}
\usepackage{physics}
\usepackage[caption=false]{subfig}
\usepackage{booktabs}
\usepackage{color}
\usepackage{comment}
\def\X{\sigma^x}
\def\Y{\sigma^y}
\def\Z{\sigma^z}
\def \ab{\bar{\alpha}}
\def \abb{\bar{\bar{\alpha}}}
\def \bb{\bar{\beta}}
\def \bbb{\bar{\bar{\beta}}}

\begin{document}

\title{Boundary Strong Zero Modes}
\author{Christopher T. Olund}
\affiliation{Department of Physics, University of California at Berkeley, Berkeley, CA 94720, USA}
\author{Norman Y. Yao}
\affiliation{Department of Physics, University of California at Berkeley, Berkeley, CA 94720, USA}
\affiliation{Department of Physics, Harvard University, MA 02138, USA}
\author{Jack Kemp}
\affiliation{Department of Physics, Harvard University, MA 02138, USA}

\date{\today}

\begin{abstract}
Strong zero modes  are edge-localized degrees of freedom capable of storing information at infinite temperature, even in systems with no disorder. 
To date, their stability has only been systematically explored at the physical edge of a system. 
Here, we extend the notion of strong zero modes to the boundary between two systems, and present a unifying framework for the stability of these boundary strong zero modes. 
Unlike zero-temperature topological edge modes, which are guaranteed to exist at the interface between a trivial and topological phase, the robustness of boundary strong zero modes is significantly more subtle.
This subtlety is perhaps best illustrated by the following dichotomy: we find that the interface between a trivial and ordered phase does not guarantee the existence of a strong zero mode, while the interface between two ordered phases can, in certain cases, lead to an \emph{exact} strong zero mode. 

\end{abstract}

\maketitle

A tremendous amount of recent interest has centered on the possibility of storing and manipulating quantum information in isolated qubits that naturally emerge from a many-body system~\cite{nayak_non-abelian_2008, alicea_new_2012, choi_quantum_2015, karzig_scalable_2017, bomantara_quantum_2018}.
Perhaps the most familiar example of this strategy is in the context of topological physics, where  the boundary of a many-body system can exhibit edge modes that are decoupled from the bulk~\cite{affleck_rigorous_1987, haldane_continuum_1983, Kitaev_2001, chen_symmetry_2013}.
However, at temperatures above the topological gap, interactions with thermally-excited quasi-particles can cause rapid decoherence ~\cite{KITAEV20032,RevModPhys.80.1083,DasSarmaTQC}. 
One approach to combat the effects of temperature is to utilize strong disorder to drive such a system into the many-body localized phase~\cite{Chandran14, Potter15, Bahri15, Yao15}. 

Interestingly, seminal recent results over the last decade have introduced an alternate strategy~\cite{Fendley12, Jermyn:2014, aasen_topological_2016, Alicea2016, Fendley_2016,Else17, Muller:2016, Kemp_2017, Moran17,Vasiloiu18, Vasiloiu19, Yates19, kemp2019symmetry, Rakovszky20, yeh_decay_2023}, whereby even translationally-invariant systems can host stable boundary degrees of freedom at finite temperature. 
In these systems, the edge is unable to resonantly absorb or emit bulk excitations. 
Such long-lived edge modes across the entire spectrum are known as `strong edge zero modes' (SZMs)~\cite{Fendley12, Alicea2016, topologynote}.
They are separated into `exact' SZMs, which decouple the boundary completely from the bulk in the thermodynamic limit, and `almost' SZMs, where the system eventually does thermalize, but parametrically slowly~\cite{Abanin17,PhysRevX.7.011026, Else17, Kemp_2017, Yates20a, Yates20b}.

In the topological setting, it is common for boundary modes to be constructed not only on the edge of a system, but also at the interface between topological and trivial phases. 
Protocols can then be devised to easily manipulate qubits by controlling the location of the phase boundary~\cite{alicea:2011, Yao15}. 
Almost by definition, topological edge modes ought to be stable at both \emph{system} boundaries and \emph{phase} boundaries.
But the equivalent question for SZMs is significantly more subtle. 
Indeed, one cannot rely on topological arguments at arbitrary temperatures, so the stability of SZMs at the boundary between two phases is not at all manifest. 
The situation is even less clear for SZMs in systems without underlying topological edge modes. 

In this Letter, we present a unifying framework for the stability of boundary strong zero modes (Table~\ref{tab:scenariotable}).
Our main results are three-fold.
First, we begin by studying the one-dimensional transverse-field Ising model (TFIM) and its fermionic equivalent, the Kitaev chain. 
For the latter, the SZM remains effectively unchanged at the boundary between a trivial and topological phase compared to the edge.
 In contrast,  for the former, a phase boundary does not guarantee the existence of a SZM. 
 We show that this difference is a natural consequence of the nonlocality of the Jordan--Wigner transformation that connects the two models.

\begin{table}[]

\includegraphics[width=0.9\linewidth]{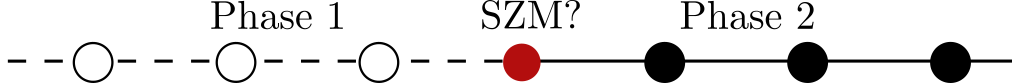}
\vspace{2pt}

\begin{tabular}{llccl}\toprule
Phase 1 & Phase 2 & SZM & Exact & Resonances \\\midrule
\textit{Kitaev chain} &&&&\\
Vacuum&Topological & \checkmark & \checkmark & None                                \\
Trivial&Topological & \checkmark & \checkmark & None                                  \\
Topological&Topological      & $\times$ & $\times$ & |                                \\
\midrule
\multicolumn{2}{l}{\textit{Transverse-field Ising model}} &&&\\
Vacuum&Ferromagnetic      & \checkmark & \checkmark & None   \\
Paramagnetic&Ferromagnetic        & \checkmark & \checkmark & Single\\
Ferromagnetic&Ferromagnetic      & \checkmark & \checkmark & Single   \\\midrule
\multicolumn{2}{l}{\textit{TFIM with interactions}}  &&&\\
Any & Ferromagnetic & \checkmark & $\times$ & Multiple\\\midrule
\multicolumn{2}{l}{\textit{XXZ chain}}  &&&\\
Vacuum & Ferromagnetic & \checkmark & \checkmark & None   \\ 
Ferromagnetic & Ferromagnetic & \checkmark & ? & Single\\
XY & Ferromagnetic & $\checkmark$ & ? & None\\
 \bottomrule
\end{tabular}

\caption{Summary of strong zero mode properties on various system edges and boundaries.}
\label{tab:scenariotable}
\end{table}

 Second, we demonstrate that the extinction of the strong zero mode in the TFIM is related to a resonance that allow bulk excitations to pass through the boundary between the ferromagnet and the paramagnet.
 Away from this resonance, we prove a surprising result, namely, that the boundary zero mode is actually an
\emph{exact} SZM. 
This represents the first example of an exact strong zero mode where the SZM expansion exhibits non-trivial poles~\cite{Fendley12, Kemp_2017}. 
Our resonance-based understanding suggests another intriguing possibility---that even the boundary between two ferromagnetic regions with different Ising couplings can exhibit an SZM. We show that this is indeed the case. 

Finally, turning on interactions, we provide both numerical and analytic evidence that, much as at system boundaries, integrability-breaking allows for resonances to proliferate and demotes the exact SZM to an almost SZM. 
On the other hand, in the interacting (but integrable) XXZ model, we conjecture that the boundary SZM is exact since it exhibits only a single resonance.

\textit{Boundary strong zero mode in the transverse field Ising model}---We begin by detailing the basic physical consequences of an edge SZM, to set the stage for our study of SZMs at the boundary between different phases. An exact strong zero mode is a quasilocal operator, $\Psi$, which commutes with the Hamiltonian up to an error exponentially small in system size,  anticommutes with a discrete symmetry, $\mathcal{F}$, of the model, and squares to the identity, $\Psi^2=1$~\cite{Kemp_2017, parafermnote}.
Physically, the existence of a SZM implies that observables with large overlap with the SZM will evolve slowly; i.e. they will retain memory of their initial state for anomalously long times.

\begin{figure} []
\includegraphics[width=0.8\linewidth]{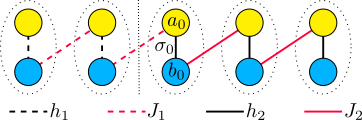}
\caption{Schematic of the boundary Ising model. The ovals denote the spins and the circles the constituent Majorana fermions.} 
 \label{fig:diagram}
\end{figure}

The simplest  example of an exact SZM occurs at the edge of the 1D transverse field Ising model (TFIM), or equivalently the free-fermionic Kitaev chain. In order to consider the boundary between phases we will couple the ends of two chains with different coupling constants together:
\vspace{-4mm}
\begin{equation}
\begin{split}
    H_\text{BI}=&-J_1\sum_{j=-\infty}^{-1}\sigma_j^z\sigma_{j+1}^z
    -J_2\sum_{j=0}^{\infty}\sigma_j^z\sigma_{j+1}^z\\
    &-h_1\sum_{j=-\infty}^{-1}\sigma_j^x-h_2\sum_{j=0}^\infty\sigma_j^x.
\end{split}
\end{equation}
Here $J_{1(2)}$ and $h_{1(2)}$ are the Ising coupling and transverse field on the left- (right-) hand chain of the system, respectively. The two chains are coupled by $J_1$ at site $j=0$ (Fig.~\ref{fig:diagram}). 

This model can also be transformed into two end-to-end coupled Kitaev chains using the standard, non-local Jordan--Wigner (JW) transformation by defining Majorana fermion operators~\cite{JW1928,Fendley_2014}:
$
    a_j= \left(\prod_{k=-\infty}^{j-1}\sigma_k^x\right)\sigma_j^z$ 
    and $
    b_j= i\left(\prod_{k=-\infty}^j\sigma_k^x\right)\sigma_j^z$.
    This yields the corresponding free-fermion Hamiltonian 
\begin{equation}
\begin{split}
    H_\text{FBI}=&-iJ_1\sum_{j=-\infty}^{-1}b_ja_{j+1}
    -iJ_2\sum_{j=0}^{\infty}b_ja_{j+1}\\
    &-ih_1\sum_{j=-\infty}^{-1}a_jb_j-ih_2\sum_{j=0}^\infty a_jb_j.
\end{split}
\end{equation}
Focusing  on the uncoupled case $J_1=0$, the second chain supports a familiar Majorana zero-mode if $J > h$~\cite{Kitaev_2001}. 
Interestingly,  the edge supports not only this famous topological edge mode in the ground state, but furthermore an exact SZM acting on the entire spectrum. 
This edge SZM is given by $a_0$ with an exponential tail into the bulk~\cite{Fendley12}. 

It is well-known that \emph{topological} edge modes (e.g. the Majorana zero-mode) exist on the boundary between trivial and topological phases, not just at the edge of a system. 
For example, in our system, the topological edge mode survives if one turns on the $J_1$ coupling between chains so long as the first chain remains in the trivial phase, $J_1 < h_1$. 
By constructing the SZM perturbatively with respect to the small couplings $J_1$ and $h_2$, from the usual zeroth order ansatz $\Psi_\text{FBI}^{(0)}=a_0$~\cite{Fendley12}, we find that an exact SZM also exists at the boundary:
\begin{equation*}
    \Psi_\text{FBI}=\mathcal{N}_\text{FBI}\left(a_0+\sum_{j=1}^\infty
    \left(\left(\frac{J_1}{h_1}\right)^ja_{-j}+
    \left(\frac{h_2}{J_2}\right)^ja_j\right)\right) \label{eq:szmfreefermion}
\end{equation*}
with \begin{equation*}
    \mathcal{N}_\text{FBI}^2=\frac{(1-(J_1/h_1)^2)(1-(h_2/J_2)^2)}
    {1-(J_1h_2/J_2h_1)^2}.
  \end{equation*}
   $\Psi_\text{FBI}$ is the same as the exact SZM of the uncoupled  chain, but is now dressed by an additional tail into the trivial phase.
   The physics of the edge and boundary SZMs  are thus essentially the same.

  We stress that this result does not immediately follow from the existence of the topological edge mode at the boundary. The topological physics occurs at energy densities below the gap, while for the SZM, we are interested in infinite temperature behavior.

  In this boundary model, the SZM does not start at the edge of the chain, but at the center, at site $0$. 
  The operator in the spin model corresponding to $a_0$  is not simply $\sigma_0^z$, but is instead the highly nonlocal operator $\left(\prod_{k=-\infty}^{-1}\sigma_k^x\right)\sigma_0^z$, as it picks up a Jordan--Wigner string.
  Thus, the existence of $\Psi_\text{FBI}$ does not imply the existence of an SZM at a phase boundary in the TFIM. This is true regardless of how we choose to define our Jordan--Wigner strings~\cite{suppinfo}.

In order to elucidate the difference between spins and Majorana fermions, we again calculate the boundary zero mode perturbatively, but starting from the boundary spin $\sigma^z_0$ as our zeroth order estimate for the purported SZM. 
This time, the number of terms grows rapidly as a function of the order $n$ in perturbation theory. 
The first two corrections beyond the zeroth order term $\Psi^{(0)}_\text{BI}=\Z_0$ are
\vspace{-2mm}
\begin{align}
    \Psi^{(1)}_\text{BI}=&\frac{h_2}{J_2}\X_0\Z_1 \\
    \Psi^{(2)}_\text{BI}=&\frac{h_2^2}{J_2^2}\X_0\X_1\Z_2
    -\frac{J_1h_2}{J_2^2-h_1^2}\Z_{-1}\X_0 \nonumber\\
    &+\frac{J_1h_1h_2}{J_2(J_2^2-h_1^2)}\Y_{-1}\Y_0\Z_1
    -\frac{h_2^2}{2J_2^2}\Z_0. \label{eq:szmfreespin}
\end{align}

Immediately, we see a stark contrast with the fermionic SZM. Instead of the SZM breaking down at the phase transition $h_1 = J_1$, there is a pole in the expansion at $J_2 = \pm h_1$. These divergent terms in the perturbation theory are called `resonances', and are caused by energy-conserving processes which change the boundary degrees of freedom~\cite{Kemp_2017}. For example, the poles at $h_1=\pm J_2$ appearing at second order describe the process in which a spin-flip excitation of the paramagnetic chain converts resonantly into a domain-wall excitation of the ferromagnetic chain.

The appearance of such resonances typically signal that the perturbative expansion of the SZM no longer converges: instead, we are forced to truncate the series before divergent terms appear. 
In this scenario, the strong zero mode is known as an `almost', rather than exact, SZM. 
In other models with known almost SZMs, such as the Ising model with  next-nearest-neighbor interactions~\cite{Kemp_2017}, parafermions~\cite{Fendley12, Jermyn:2014, Moran17} and the ZXZ model~\cite{kemp2019symmetry}, there are an increasing number of different poles as we go to higher orders in $\Psi$, representing more and more complicated resonant processes that only appear at higher order.

In our boundary Ising model, however, there is a crucial difference: no further poles emerge. Poles at all higher orders are still at $h_1=\pm J_2$. This can be intuitively understood as the integrability of the model preventing any resonant processes except for the conversion of a single domain wall into a single spin-flip. For example, naively our argument above suggests a pole should exist at $h_1=\pm 2 J_2$, because we could convert \emph{two} domain walls resonantly into a single spin flip at the boundary. In a non-integrable model, this argument would be correct; however, it is not possible to construct such a three-body resonant conversion at the boundary in an integrable model without it factoring into a pair of two-body processes, one of which does not involve the boundary degree of freedom~\cite{SML}.

As there is only a single resonance, it is still possible that the SZM expansion converges.  
Furthermore, the lack of dependence on the phase transition at $h_1=J_1$ suggests repeating the calculation above for $J_1 > h_1$.  This reveals the SZM is present at the boundary of two ferromagnetic regions as  well, with a pole at $J_1=\pm J_2$. Thus for the spin boundary model a phase boundary is not even a prerequisite for a SZM---indeed, the left-hand chain can even be critical, $J_1=h_1$.

\begin{figure} []
\includegraphics[width=0.9\linewidth]{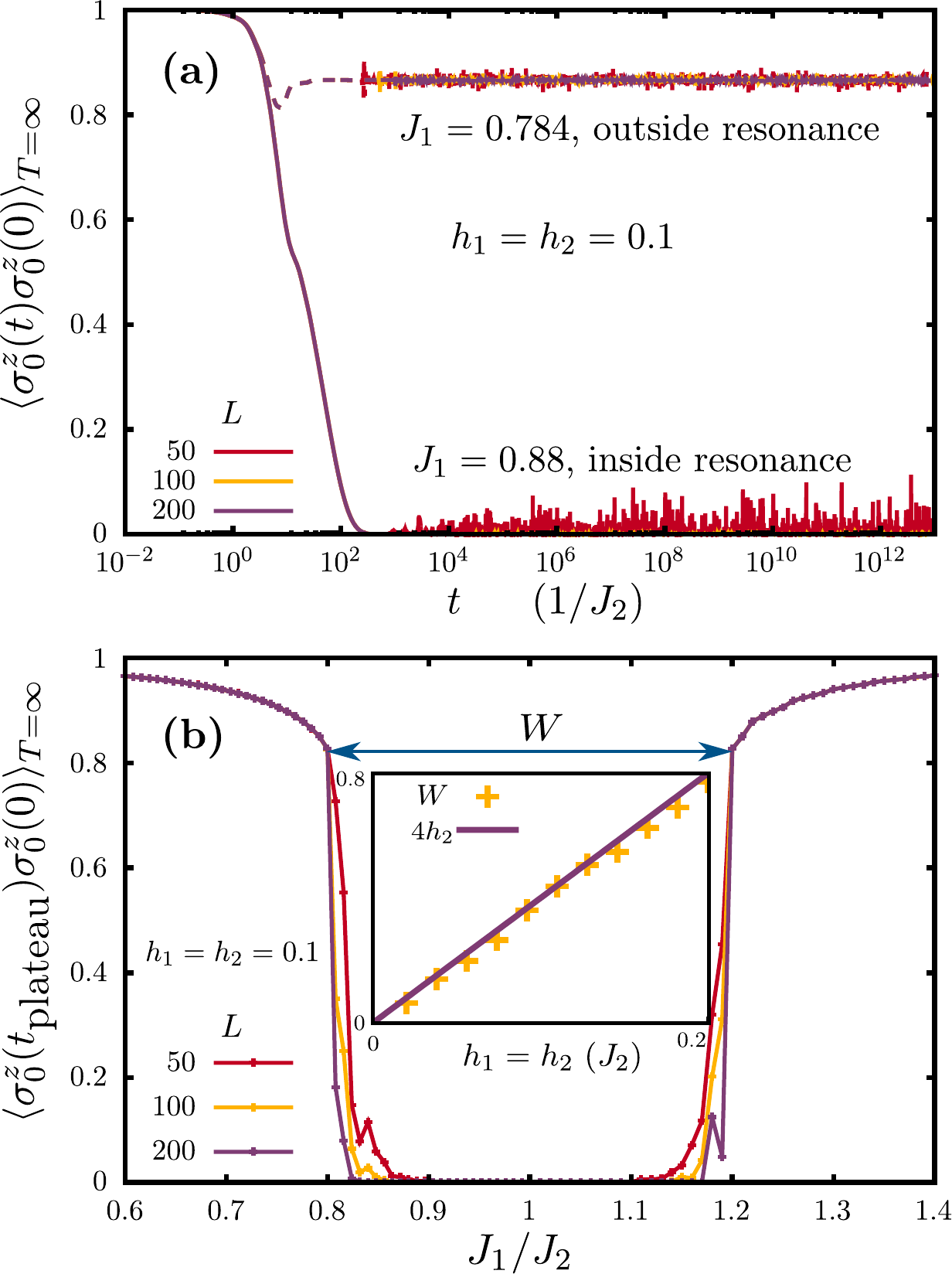}
\caption{\textbf{(a)} The  autocorrelator for the spin at the boundary of the boundary transverse-field Ising model for various systems sizes $L$ for two values of $J_1$ outside and inside the resonance. \textbf{(b)} The plateau value of the autocorrelator for the spin at the boundary of the boundary transverse-field Ising model for various systems sizes $L$ as the coupling $h_1$ is varied, for $h_1 = h_2 = 0.1 J_2$. Nearby $J_1 = J_2$ a resonance suppresses the SZM. \textbf{Inset:} The width of the resonance at $J_1 = J_2$, as a function of $h_1=h_2$. This is estimated from the value of $J_1$ for which the plateau value of the autocorrelator falls below 0.8 or above 1.2.}
 \label{fig:spinautocorr}
\end{figure}

\textit{Physical Consequences}---A direct physical consequence of the SZM is reflected in the infinite temperature autocorrelator of the boundary spin magnetization $A_\infty(t)=\langle\sigma_0^z(t)\sigma_0^z(0)\rangle_{T=\infty}$.

If the system supports an exact SZM with finite overlap with the boundary spin, then  the decay time of this autocorrelator will be infinite in the thermodynamic limit.

We can calculate the autocorrelation time of the boundary spin $\sigma^z_0$ numerically from the free fermion results \cite{bravyi:2012}. In Fig.~\ref{fig:spinautocorr}a), we plot the autocorrelator for the spin at the ferromagnetic-ferromagnetic boundary. Outside the resonance, the autocorrelation time increases exponentially with system size~\cite{oscillatenote}. In stark contrast, close to the resonance the autocorrelator immediately decays.

We plot the plateau value of the $\sigma^z_0$ autocorrelator in Fig.~\ref{fig:spinautocorr}b). If the SZM is exact, this is the asymptotic value of the autocorrelator in the thermodynamic limit. Unlike for the fermions, we observe that the location of the phase transition is unimportant: rather, the amplitude dramatically decreases around the resonance. Crucially, the width of the resonance does not seem to increase with system size, suggesting that it remains finite in the thermodynamic limit, and outside the resonance the SZM is exact.

Intuitively, the SZM suffers a resonance when excitations in the left-hand chain can move into the right-hand chain without any change in the overall energy of the system. If we neglect the connection between the chains, we may calculate their individual energy spectra exactly~\cite{SML}. For the range of couplings considered in Fig~\ref{fig:spinautocorr}b), the bands of the two chains overlap when $J_2-2 h_2< J_1< J_2 + 2 h_2$. This picture is consistent with the growth of the width of the resonance in the autocorrelation time observed in the inset of Fig.~\ref{fig:spinautocorr}b).

Analytically, if the SZM is exact, the operator expansion should converge to all orders in perturbation theory. We present three results in the supplemental material that show this indeed must be the case~\cite{suppinfo}. Firstly, using simple combinatorial arguments, we bound the number of terms that first appear at order $n$ in perturbation theory to a subexponential in $n$. Secondly, we solve exactly a simple toy model where the semi-infinite chain on the left of the boundary is replaced by a single spin. The solution is valid for all choices of the couplings $J_1$ and $h_1$ on the left-hand chain, unless proximity to the resonance prohibits convergence. We confirm that the normalization extracted from this solution agrees precisely with numerics~\cite{suppinfo}. The boundary Ising model with a single spin at the edge, is, therefore, an exact SZM within the radius of convergence, despite the existence of a pole.

Although the full solution is quite complicated and involves powers of the generating function for the Narayana numbers~\cite{petersen:2015}, for a bit of intuition, it is instructive to consider the  simple case, $h_1=0$. The solution for the SZM reduces to:
\vspace{-2mm}
\begin{align*}\Psi_2=&\mathcal{N}_2\sum^\infty_{\mathclap{n, m=-1}} C_{nm}\sigma^x_{-1}\left(\prod_{k=-1}^{n}\sigma_k^x\right)\sigma_{n+1}^z \left(\prod_{k=-1}^m\sigma_k^x\right)\sigma_m^z \\[10pt]
C_{nm} &= \left(\frac{J_1}{J_2}\right)^{\mathbf{1}_{m>-1}}\left(\frac{h_2J_2}{J_2^2-J_1^2}\right)^{n+1} \left(\frac{h_2J_2}{J_2^2-J_1^2}\right)^{m+1},
\end{align*}
where $\mathbf{1}_{m>-1} = 1$ if $m>-1$ and otherwise vanishes. It is evident $\Psi_2$ is normalizable so long as $|h_2J_2/(J_2^2-J_1^2)|<1$, with a normalization factor:
\begin{equation}
    \mathcal{N}_2=\frac{1}{\sqrt{1-\frac{h_2^2}{J_2^2-J_1^2}}}\left(1-\left(\frac{h_2J_2}{J_2^2-J_1^2}\right)^2\right).
  \end{equation}

Finally, guided by the exact solution to the toy model, we show that for the general case with $N$ spins to the left of the boundary, all terms at order $n$ have values that are exponentially suppressed in $n$.
Thus, the SZM is both localized at the boundary, and has a finite normalization.
In the limit $N \to \infty$, the bounds on this convergence are:
\vspace{-3mm}
\begin{equation}
\left|\frac{(J_2^2-J_1^2+h_2^2-h_1^2)\pm 2 h_1 J_1}{h_2 J_2} \right| >2, \qquad \left|\frac{h_2}{J_2}\right|< 1,
\end{equation}
which are both consistent with our simple physical argument based on band overlaps, and the numerics presented in Fig.~\ref{fig:spinautocorr}b).

\textit{Interactions}---The systems we have so far considered constitute a special case, as they are not only integrable, but free. We now consider adding interactions. We have two possibilites: most naively, we can add integrability breaking interactions, but we can also consider integrable but interacting systems. To study a non-integrable, interacting system we add nearest-neighbor XX interactions as a perturbation to our coupled TFIM chains of the form $-K\sum_{j=-\infty}^\infty \X_j\X_{j+1}$, with $K\ll J_2$.

With these terms added, both spin and fermion models now exhibit an almost SZM rather than an exact SZM, with additional resonances appearing at higher order in perturbation theory. For example, in addition to resonances at $J_2=\pm J_1$, the SZM also has poles starting at third order at $J_2=\pm (3)^{\pm 1} J_1$, and generically at order $p+q-1$ for $J_2 = \pm(p/q)^{\pm 1} J_1$ for $p,q$ odd integers.

\begin{figure} []

 \includegraphics[width=1\columnwidth]{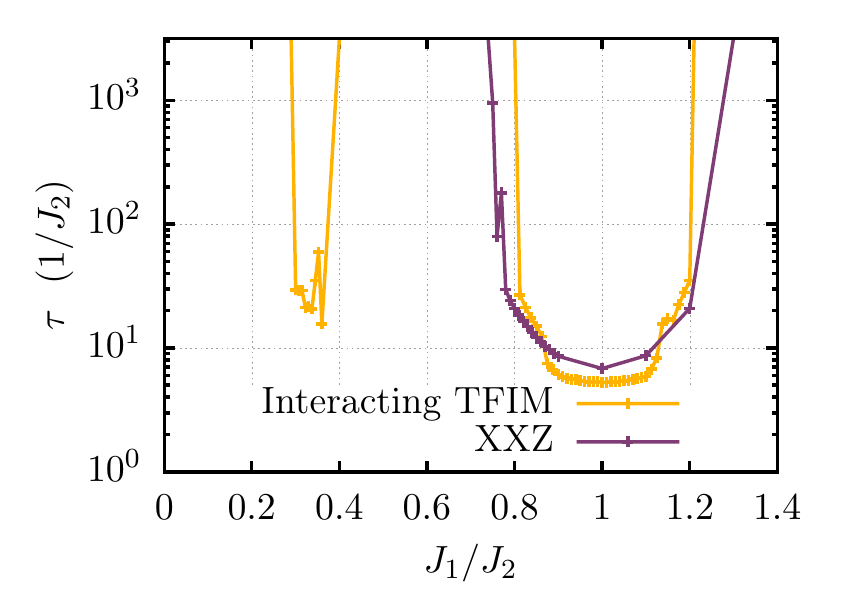}

\caption{The decay time for the boundary spin $\sigma^z_0$ between ferromagnetic and paramagnetic regimes in the presence of interactions for system size $L = 24$ and $h_1=h_2 = 0.1 J_2$ from Krylov subspace methods~\cite{meyer:2020}. Other than at the poles at $J_1$ = $J_2$ and $J_2/3$ the decay time is longer than the evolution time.}
\label{fig:interactions}

\end{figure}

Numerically, we can extract the decay time from when the autocorrelator falls below a threshold value, Fig.~\ref{fig:interactions}. Even for large values of the interaction strengths, we see that the boundary degrees of freedom have significantly enhanced lifetimes. Additionally, the effect of the resonances can be seen as dips in the autocorrelation times exactly at the values predicted above.

For interacting, integrable systems, we do not expect this proliferation of resonances. For example, consider coupling two XXZ chains end-to-end together as we coupled the two TFIM chains:
\vspace{-2mm}
\begin{align*}
    H_\text{XXZ}=&-\sum_{j=-\infty}^{-1}\left[J_1 \sigma^z_j \sigma^z_{j+1}+ h_1( \sigma^x_j\sigma^x_{j+1}+\sigma^y_j\sigma^y_{j+1}) \right]\\
    &- \sum_{j=0}^{\infty}\left[J_2 \sigma^z_j \sigma^z_{j+1}+ h_2( \sigma^x_j\sigma^x_{j+1}+\sigma^y_j\sigma^y_{j+1}) \right].
\end{align*}

As for the TFIM, we will set the second chain to always be ferromagnetic, $J_2 > h_2$. If $J_1 < h_1$, then the first chain is gapless, and there are no resonances: the SZM exists if and only if $h_1$ is sufficiently smaller than $J_2$. On the other hand, if both chains are ferromagnetic, there is a single resonance at $J_1 = J_2$ (Fig.~\ref{fig:interactions}).

\textit{Outlook}---Our work opens the door to a number of intriguing directions.  Most directly, is it possible to calculate an explicit solution for the exact boundary SZM, which appears to have an intriguing relation to squares of the Narayana numbers \cite{suppinfo}? Furthermore, is the boundary SZM also exact for interacting integrable systems, such as the XXZ chain, which is known to host an exact SZM at a system boundary? If so, it may be possible to relate this conserved quantity to the hierarchy of conserved quantities that have recently been constructed in the XXZ chain~\cite{grabowski:1994,nozawa:2020, nienhuis:2021}. For practical purposes, it will be important to study how stable the SZMs are under dynamical protocols to manipulate them by moving the phase boundaries~\cite{Yao15}.

We gratefully acknowledge discussions with C. Laumann, F. Machado, and P. Fendley. 
This work was supported by the NSF through the QII-TAQS program (Grant No. 1936100) and the QLCI program (Grant No. OMA-2016245), and the David and Lucile Packard foundation.

\bibliography{BSZM}

\begin{thebibliography}{49}%
\makeatletter
\providecommand \@ifxundefined [1]{%
 \@ifx{#1\undefined}
}%
\providecommand \@ifnum [1]{%
 \ifnum #1\expandafter \@firstoftwo
 \else \expandafter \@secondoftwo
 \fi
}%
\providecommand \@ifx [1]{%
 \ifx #1\expandafter \@firstoftwo
 \else \expandafter \@secondoftwo
 \fi
}%
\providecommand \natexlab [1]{#1}%
\providecommand \enquote  [1]{``#1''}%
\providecommand \bibnamefont  [1]{#1}%
\providecommand \bibfnamefont [1]{#1}%
\providecommand \citenamefont [1]{#1}%
\providecommand \href@noop [0]{\@secondoftwo}%
\providecommand \href [0]{\begingroup \@sanitize@url \@href}%
\providecommand \@href[1]{\@@startlink{#1}\@@href}%
\providecommand \@@href[1]{\endgroup#1\@@endlink}%
\providecommand \@sanitize@url [0]{\catcode `\\12\catcode `\$12\catcode
  `\&12\catcode `\#12\catcode `\^12\catcode `\_12\catcode `\%12\relax}%
\providecommand \@@startlink[1]{}%
\providecommand \@@endlink[0]{}%
\providecommand \url  [0]{\begingroup\@sanitize@url \@url }%
\providecommand \@url [1]{\endgroup\@href {#1}{\urlprefix }}%
\providecommand \urlprefix  [0]{URL }%
\providecommand \Eprint [0]{\href }%
\providecommand \doibase [0]{https://doi.org/}%
\providecommand \selectlanguage [0]{\@gobble}%
\providecommand \bibinfo  [0]{\@secondoftwo}%
\providecommand \bibfield  [0]{\@secondoftwo}%
\providecommand \translation [1]{[#1]}%
\providecommand \BibitemOpen [0]{}%
\providecommand \bibitemStop [0]{}%
\providecommand \bibitemNoStop [0]{.\EOS\space}%
\providecommand \EOS [0]{\spacefactor3000\relax}%
\providecommand \BibitemShut  [1]{\csname bibitem#1\endcsname}%
\let\auto@bib@innerbib\@empty
\bibitem [{\citenamefont {Nayak}\ \emph
  {et~al.}(2008{\natexlab{a}})\citenamefont {Nayak}, \citenamefont {Simon},
  \citenamefont {Stern}, \citenamefont {Freedman},\ and\ \citenamefont
  {Das~Sarma}}]{nayak_non-abelian_2008}%
  \BibitemOpen
  \bibfield  {author} {\bibinfo {author} {\bibfnamefont {C.}~\bibnamefont
  {Nayak}}, \bibinfo {author} {\bibfnamefont {S.~H.}\ \bibnamefont {Simon}},
  \bibinfo {author} {\bibfnamefont {A.}~\bibnamefont {Stern}}, \bibinfo
  {author} {\bibfnamefont {M.}~\bibnamefont {Freedman}},\ and\ \bibinfo
  {author} {\bibfnamefont {S.}~\bibnamefont {Das~Sarma}},\ }\bibfield  {title}
  {\bibinfo {title} {Non-{Abelian} anyons and topological quantum
  computation},\ }\href {https://doi.org/10.1103/RevModPhys.80.1083} {\bibfield
   {journal} {\bibinfo  {journal} {Reviews of Modern Physics}\ }\textbf
  {\bibinfo {volume} {80}},\ \bibinfo {pages} {1083} (\bibinfo {year}
  {2008}{\natexlab{a}})}\BibitemShut {NoStop}%
\bibitem [{\citenamefont {Alicea}(2012)}]{alicea_new_2012}%
  \BibitemOpen
  \bibfield  {author} {\bibinfo {author} {\bibfnamefont {J.}~\bibnamefont
  {Alicea}},\ }\bibfield  {title} {\bibinfo {title} {New directions in the
  pursuit of {Majorana} fermions in solid state systems},\ }\href
  {https://doi.org/10.1088/0034-4885/75/7/076501} {\bibfield  {journal}
  {\bibinfo  {journal} {Reports on Progress in Physics}\ }\textbf {\bibinfo
  {volume} {75}},\ \bibinfo {pages} {076501} (\bibinfo {year}
  {2012})}\BibitemShut {NoStop}%
\bibitem [{\citenamefont {Choi}\ \emph {et~al.}(2015)\citenamefont {Choi},
  \citenamefont {Yao}, \citenamefont {Gopalakrishnan},\ and\ \citenamefont
  {Lukin}}]{choi_quantum_2015}%
  \BibitemOpen
  \bibfield  {author} {\bibinfo {author} {\bibfnamefont {S.}~\bibnamefont
  {Choi}}, \bibinfo {author} {\bibfnamefont {N.~Y.}\ \bibnamefont {Yao}},
  \bibinfo {author} {\bibfnamefont {S.}~\bibnamefont {Gopalakrishnan}},\ and\
  \bibinfo {author} {\bibfnamefont {M.~D.}\ \bibnamefont {Lukin}},\ }\bibfield
  {title} {\bibinfo {title} {Quantum {Control} of {Many}-body {Localized}
  {States}},\ }\href@noop {} {\  (\bibinfo {year} {2015})},\ \Eprint
  {https://arxiv.org/abs/1508.06992} {arXiv:1508.06992} \BibitemShut {NoStop}%
\bibitem [{\citenamefont {Karzig}\ \emph {et~al.}(2017)\citenamefont {Karzig},
  \citenamefont {Knapp}, \citenamefont {Lutchyn}, \citenamefont {Bonderson},
  \citenamefont {Hastings}, \citenamefont {Nayak}, \citenamefont {Alicea},
  \citenamefont {Flensberg}, \citenamefont {Plugge}, \citenamefont {Oreg},
  \citenamefont {Marcus},\ and\ \citenamefont
  {Freedman}}]{karzig_scalable_2017}%
  \BibitemOpen
  \bibfield  {author} {\bibinfo {author} {\bibfnamefont {T.}~\bibnamefont
  {Karzig}}, \bibinfo {author} {\bibfnamefont {C.}~\bibnamefont {Knapp}},
  \bibinfo {author} {\bibfnamefont {R.~M.}\ \bibnamefont {Lutchyn}}, \bibinfo
  {author} {\bibfnamefont {P.}~\bibnamefont {Bonderson}}, \bibinfo {author}
  {\bibfnamefont {M.~B.}\ \bibnamefont {Hastings}}, \bibinfo {author}
  {\bibfnamefont {C.}~\bibnamefont {Nayak}}, \bibinfo {author} {\bibfnamefont
  {J.}~\bibnamefont {Alicea}}, \bibinfo {author} {\bibfnamefont
  {K.}~\bibnamefont {Flensberg}}, \bibinfo {author} {\bibfnamefont
  {S.}~\bibnamefont {Plugge}}, \bibinfo {author} {\bibfnamefont
  {Y.}~\bibnamefont {Oreg}}, \bibinfo {author} {\bibfnamefont {C.~M.}\
  \bibnamefont {Marcus}},\ and\ \bibinfo {author} {\bibfnamefont {M.~H.}\
  \bibnamefont {Freedman}},\ }\bibfield  {title} {\bibinfo {title} {Scalable
  designs for quasiparticle-poisoning-protected topological quantum computation
  with {Majorana} zero modes},\ }\href
  {https://doi.org/10.1103/PhysRevB.95.235305} {\bibfield  {journal} {\bibinfo
  {journal} {Physical Review B}\ }\textbf {\bibinfo {volume} {95}},\ \bibinfo
  {pages} {235305} (\bibinfo {year} {2017})}\BibitemShut {NoStop}%
\bibitem [{\citenamefont {Bomantara}\ and\ \citenamefont
  {Gong}(2018)}]{bomantara_quantum_2018}%
  \BibitemOpen
  \bibfield  {author} {\bibinfo {author} {\bibfnamefont {R.~W.}\ \bibnamefont
  {Bomantara}}\ and\ \bibinfo {author} {\bibfnamefont {J.}~\bibnamefont
  {Gong}},\ }\bibfield  {title} {\bibinfo {title} {Quantum computation via
  {Floquet} topological edge modes},\ }\href
  {https://doi.org/10.1103/PhysRevB.98.165421} {\bibfield  {journal} {\bibinfo
  {journal} {Physical Review B}\ }\textbf {\bibinfo {volume} {98}},\ \bibinfo
  {pages} {165421} (\bibinfo {year} {2018})}\BibitemShut {NoStop}%
\bibitem [{\citenamefont {Affleck}\ \emph {et~al.}(1987)\citenamefont
  {Affleck}, \citenamefont {Kennedy}, \citenamefont {Lieb},\ and\ \citenamefont
  {Tasaki}}]{affleck_rigorous_1987}%
  \BibitemOpen
  \bibfield  {author} {\bibinfo {author} {\bibfnamefont {I.}~\bibnamefont
  {Affleck}}, \bibinfo {author} {\bibfnamefont {T.}~\bibnamefont {Kennedy}},
  \bibinfo {author} {\bibfnamefont {E.~H.}\ \bibnamefont {Lieb}},\ and\
  \bibinfo {author} {\bibfnamefont {H.}~\bibnamefont {Tasaki}},\ }\bibfield
  {title} {\bibinfo {title} {Rigorous results on valence-bond ground states in
  antiferromagnets},\ }\href {https://doi.org/10.1103/PhysRevLett.59.799}
  {\bibfield  {journal} {\bibinfo  {journal} {Physical Review Letters}\
  }\textbf {\bibinfo {volume} {59}},\ \bibinfo {pages} {799} (\bibinfo {year}
  {1987})}\BibitemShut {NoStop}%
\bibitem [{\citenamefont {Haldane}(1983)}]{haldane_continuum_1983}%
  \BibitemOpen
  \bibfield  {author} {\bibinfo {author} {\bibfnamefont {F.~D.~M.}\
  \bibnamefont {Haldane}},\ }\bibfield  {title} {\bibinfo {title} {Continuum
  dynamics of the 1-{D} {Heisenberg} antiferromagnet: {Identification} with the
  {O}(3) nonlinear sigma model},\ }\href
  {https://doi.org/10.1016/0375-9601(83)90631-X} {\bibfield  {journal}
  {\bibinfo  {journal} {Physics Letters A}\ }\textbf {\bibinfo {volume} {93}},\
  \bibinfo {pages} {464} (\bibinfo {year} {1983})}\BibitemShut {NoStop}%
\bibitem [{\citenamefont {Kitaev}(2001)}]{Kitaev_2001}%
  \BibitemOpen
  \bibfield  {author} {\bibinfo {author} {\bibfnamefont {A.~Y.}\ \bibnamefont
  {Kitaev}},\ }\bibfield  {title} {\bibinfo {title} {Unpaired majorana fermions
  in quantum wires},\ }\href {https://doi.org/10.1070/1063-7869/44/10s/s29}
  {\bibfield  {journal} {\bibinfo  {journal} {Physics-Uspekhi}\ }\textbf
  {\bibinfo {volume} {44}},\ \bibinfo {pages} {131} (\bibinfo {year}
  {2001})}\BibitemShut {NoStop}%
\bibitem [{\citenamefont {Chen}\ \emph {et~al.}(2013)\citenamefont {Chen},
  \citenamefont {Gu}, \citenamefont {Liu},\ and\ \citenamefont
  {Wen}}]{chen_symmetry_2013}%
  \BibitemOpen
  \bibfield  {author} {\bibinfo {author} {\bibfnamefont {X.}~\bibnamefont
  {Chen}}, \bibinfo {author} {\bibfnamefont {Z.-C.}\ \bibnamefont {Gu}},
  \bibinfo {author} {\bibfnamefont {Z.-X.}\ \bibnamefont {Liu}},\ and\ \bibinfo
  {author} {\bibfnamefont {X.-G.}\ \bibnamefont {Wen}},\ }\bibfield  {title}
  {\bibinfo {title} {Symmetry protected topological orders and the group
  cohomology of their symmetry group},\ }\href
  {https://doi.org/10.1103/PhysRevB.87.155114} {\bibfield  {journal} {\bibinfo
  {journal} {Physical Review B}\ }\textbf {\bibinfo {volume} {87}},\ \bibinfo
  {pages} {155114} (\bibinfo {year} {2013})}\BibitemShut {NoStop}%
\bibitem [{\citenamefont {Kitaev}(2003)}]{KITAEV20032}%
  \BibitemOpen
  \bibfield  {author} {\bibinfo {author} {\bibfnamefont {A.}~\bibnamefont
  {Kitaev}},\ }\bibfield  {title} {\bibinfo {title} {Fault-tolerant quantum
  computation by anyons},\ }\href
  {https://doi.org/https://doi.org/10.1016/S0003-4916(02)00018-0} {\bibfield
  {journal} {\bibinfo  {journal} {Annals of Physics}\ }\textbf {\bibinfo
  {volume} {303}},\ \bibinfo {pages} {2 } (\bibinfo {year} {2003})}\BibitemShut
  {NoStop}%
\bibitem [{\citenamefont {Nayak}\ \emph
  {et~al.}(2008{\natexlab{b}})\citenamefont {Nayak}, \citenamefont {Simon},
  \citenamefont {Stern}, \citenamefont {Freedman},\ and\ \citenamefont
  {Das~Sarma}}]{RevModPhys.80.1083}%
  \BibitemOpen
  \bibfield  {author} {\bibinfo {author} {\bibfnamefont {C.}~\bibnamefont
  {Nayak}}, \bibinfo {author} {\bibfnamefont {S.~H.}\ \bibnamefont {Simon}},
  \bibinfo {author} {\bibfnamefont {A.}~\bibnamefont {Stern}}, \bibinfo
  {author} {\bibfnamefont {M.}~\bibnamefont {Freedman}},\ and\ \bibinfo
  {author} {\bibfnamefont {S.}~\bibnamefont {Das~Sarma}},\ }\bibfield  {title}
  {\bibinfo {title} {Non-abelian anyons and topological quantum computation},\
  }\href {https://doi.org/10.1103/RevModPhys.80.1083} {\bibfield  {journal}
  {\bibinfo  {journal} {Rev. Mod. Phys.}\ }\textbf {\bibinfo {volume} {80}},\
  \bibinfo {pages} {1083} (\bibinfo {year} {2008}{\natexlab{b}})}\BibitemShut
  {NoStop}%
\bibitem [{\citenamefont {Sarma}\ \emph {et~al.}(2015)\citenamefont {Sarma},
  \citenamefont {Freedman},\ and\ \citenamefont {Nayak}}]{DasSarmaTQC}%
  \BibitemOpen
  \bibfield  {author} {\bibinfo {author} {\bibfnamefont {S.~D.}\ \bibnamefont
  {Sarma}}, \bibinfo {author} {\bibfnamefont {M.}~\bibnamefont {Freedman}},\
  and\ \bibinfo {author} {\bibfnamefont {C.}~\bibnamefont {Nayak}},\ }\bibfield
   {title} {\bibinfo {title} {Majorana zero modes and topological quantum
  computation},\ }\href {https://doi.org/10.1038/npjqi.2015.1} {\bibfield
  {journal} {\bibinfo  {journal} {npj Quantum Information}\ }\textbf {\bibinfo
  {volume} {1}},\ \bibinfo {pages} {15001} (\bibinfo {year}
  {2015})}\BibitemShut {NoStop}%
\bibitem [{\citenamefont {Chandran}\ \emph {et~al.}(2014)\citenamefont
  {Chandran}, \citenamefont {Khemani}, \citenamefont {Laumann},\ and\
  \citenamefont {Sondhi}}]{Chandran14}%
  \BibitemOpen
  \bibfield  {author} {\bibinfo {author} {\bibfnamefont {A.}~\bibnamefont
  {Chandran}}, \bibinfo {author} {\bibfnamefont {V.}~\bibnamefont {Khemani}},
  \bibinfo {author} {\bibfnamefont {C.~R.}\ \bibnamefont {Laumann}},\ and\
  \bibinfo {author} {\bibfnamefont {S.~L.}\ \bibnamefont {Sondhi}},\ }\bibfield
   {title} {\bibinfo {title} {Many-body localization and symmetry-protected
  topological order},\ }\href {https://doi.org/10.1103/PhysRevB.89.144201}
  {\bibfield  {journal} {\bibinfo  {journal} {Phys. Rev. B}\ }\textbf {\bibinfo
  {volume} {89}},\ \bibinfo {pages} {144201} (\bibinfo {year}
  {2014})}\BibitemShut {NoStop}%
\bibitem [{\citenamefont {Potter}\ and\ \citenamefont
  {Vishwanath}(2015)}]{Potter15}%
  \BibitemOpen
  \bibfield  {author} {\bibinfo {author} {\bibfnamefont {A.~C.}\ \bibnamefont
  {Potter}}\ and\ \bibinfo {author} {\bibfnamefont {A.}~\bibnamefont
  {Vishwanath}},\ }\bibfield  {title} {\bibinfo {title} {Protection of
  topological order by symmetry and many-body localization},\ }\href@noop {} {\
   (\bibinfo {year} {2015})},\ \Eprint {https://arxiv.org/abs/1506.00592}
  {arXiv:1506.00592 [cond-mat.dis-nn]} \BibitemShut {NoStop}%
\bibitem [{\citenamefont {Bahri}\ \emph {et~al.}(2015)\citenamefont {Bahri},
  \citenamefont {Vosk}, \citenamefont {Altman},\ and\ \citenamefont
  {Vishwanath}}]{Bahri15}%
  \BibitemOpen
  \bibfield  {author} {\bibinfo {author} {\bibfnamefont {Y.}~\bibnamefont
  {Bahri}}, \bibinfo {author} {\bibfnamefont {R.}~\bibnamefont {Vosk}},
  \bibinfo {author} {\bibfnamefont {E.}~\bibnamefont {Altman}},\ and\ \bibinfo
  {author} {\bibfnamefont {A.}~\bibnamefont {Vishwanath}},\ }\bibfield  {title}
  {\bibinfo {title} {Localization and topology protected quantum coherence at
  the edge of hot matter},\ }\href {http://dx.doi.org/10.1038/ncomms8341}
  {\bibfield  {journal} {\bibinfo  {journal} {Nature Communications}\ }\textbf
  {\bibinfo {volume} {6}},\ \bibinfo {pages} {7341 EP } (\bibinfo {year}
  {2015})},\ \Eprint {https://arxiv.org/abs/1307.4092} {1307.4092} \BibitemShut
  {NoStop}%
\bibitem [{\citenamefont {Yao}\ \emph {et~al.}(2015)\citenamefont {Yao},
  \citenamefont {Laumann},\ and\ \citenamefont {Vishwanath}}]{Yao15}%
  \BibitemOpen
  \bibfield  {author} {\bibinfo {author} {\bibfnamefont {N.~Y.}\ \bibnamefont
  {Yao}}, \bibinfo {author} {\bibfnamefont {C.~R.}\ \bibnamefont {Laumann}},\
  and\ \bibinfo {author} {\bibfnamefont {A.}~\bibnamefont {Vishwanath}},\
  }\href@noop {} {\bibinfo {title} {Many-body localization protected quantum
  state transfer}} (\bibinfo {year} {2015}),\ \Eprint
  {https://arxiv.org/abs/1508.06995} {arXiv:1508.06995 [quant-ph]} \BibitemShut
  {NoStop}%
\bibitem [{\citenamefont {{Fendley}}(2012)}]{Fendley12}%
  \BibitemOpen
  \bibfield  {author} {\bibinfo {author} {\bibfnamefont {P.}~\bibnamefont
  {{Fendley}}},\ }\bibfield  {title} {\bibinfo {title} {{Parafermionic edge
  zero modes in Z$_{n}$-invariant spin chains}},\ }\href
  {https://doi.org/10.1088/1742-5468/2012/11/P11020} {\bibfield  {journal}
  {\bibinfo  {journal} {J.~Stat.~Mech.}\ }\textbf {\bibinfo {volume} {11}},\
  \bibinfo {pages} {20} (\bibinfo {year} {2012})}\BibitemShut {NoStop}%
\bibitem [{\citenamefont {Jermyn}\ \emph {et~al.}(2014)\citenamefont {Jermyn},
  \citenamefont {Mong}, \citenamefont {Alicea},\ and\ \citenamefont
  {Fendley}}]{Jermyn:2014}%
  \BibitemOpen
  \bibfield  {author} {\bibinfo {author} {\bibfnamefont {A.}~\bibnamefont
  {Jermyn}}, \bibinfo {author} {\bibfnamefont {R.}~\bibnamefont {Mong}},
  \bibinfo {author} {\bibfnamefont {J.}~\bibnamefont {Alicea}},\ and\ \bibinfo
  {author} {\bibfnamefont {P.}~\bibnamefont {Fendley}},\ }\bibfield  {title}
  {\bibinfo {title} {Stability of zero modes in parafermion chains},\ }\href
  {https://doi.org/10.1103/PhysRevB.90.165106} {\bibfield  {journal} {\bibinfo
  {journal} {Phys. Rev. B}\ }\textbf {\bibinfo {volume} {90}},\ \bibinfo
  {pages} {165106} (\bibinfo {year} {2014})}\BibitemShut {NoStop}%
\bibitem [{\citenamefont {Aasen}\ \emph {et~al.}(2016)\citenamefont {Aasen},
  \citenamefont {Mong},\ and\ \citenamefont
  {Fendley}}]{aasen_topological_2016}%
  \BibitemOpen
  \bibfield  {author} {\bibinfo {author} {\bibfnamefont {D.}~\bibnamefont
  {Aasen}}, \bibinfo {author} {\bibfnamefont {R.~S.~K.}\ \bibnamefont {Mong}},\
  and\ \bibinfo {author} {\bibfnamefont {P.}~\bibnamefont {Fendley}},\
  }\bibfield  {title} {\bibinfo {title} {Topological defects on the lattice:
  {I}. {The} {Ising} model},\ }\href
  {https://doi.org/10.1088/1751-8113/49/35/354001} {\bibfield  {journal}
  {\bibinfo  {journal} {Journal of Physics A: Mathematical and Theoretical}\
  }\textbf {\bibinfo {volume} {49}},\ \bibinfo {pages} {354001} (\bibinfo
  {year} {2016})}\BibitemShut {NoStop}%
\bibitem [{\citenamefont {Alicea}\ and\ \citenamefont
  {Fendley}(2016)}]{Alicea2016}%
  \BibitemOpen
  \bibfield  {author} {\bibinfo {author} {\bibfnamefont {J.}~\bibnamefont
  {Alicea}}\ and\ \bibinfo {author} {\bibfnamefont {P.}~\bibnamefont
  {Fendley}},\ }\bibfield  {title} {\bibinfo {title} {Topological phases with
  parafermions: Theory and blueprints},\ }\href
  {https://doi.org/10.1146/annurev-conmatphys-031115-011336} {\bibfield
  {journal} {\bibinfo  {journal} {Annual Review of Condensed Matter Physics}\
  }\textbf {\bibinfo {volume} {7}},\ \bibinfo {pages} {119} (\bibinfo {year}
  {2016})},\ \Eprint
  {https://arxiv.org/abs/https://doi.org/10.1146/annurev-conmatphys-031115-011336}
  {https://doi.org/10.1146/annurev-conmatphys-031115-011336} \BibitemShut
  {NoStop}%
\bibitem [{\citenamefont {Fendley}(2016)}]{Fendley_2016}%
  \BibitemOpen
  \bibfield  {author} {\bibinfo {author} {\bibfnamefont {P.}~\bibnamefont
  {Fendley}},\ }\bibfield  {title} {\bibinfo {title} {Strong zero modes and
  eigenstate phase transitions in the {XYZ}/interacting majorana chain},\
  }\href {https://doi.org/10.1088/1751-8113/49/30/30lt01} {\bibfield  {journal}
  {\bibinfo  {journal} {Journal of Physics A: Mathematical and Theoretical}\
  }\textbf {\bibinfo {volume} {49}},\ \bibinfo {pages} {30LT01} (\bibinfo
  {year} {2016})}\BibitemShut {NoStop}%
\bibitem [{\citenamefont {Else}\ \emph
  {et~al.}(2017{\natexlab{a}})\citenamefont {Else}, \citenamefont {Fendley},
  \citenamefont {Kemp},\ and\ \citenamefont {Nayak}}]{Else17}%
  \BibitemOpen
  \bibfield  {author} {\bibinfo {author} {\bibfnamefont {D.~V.}\ \bibnamefont
  {Else}}, \bibinfo {author} {\bibfnamefont {P.}~\bibnamefont {Fendley}},
  \bibinfo {author} {\bibfnamefont {J.}~\bibnamefont {Kemp}},\ and\ \bibinfo
  {author} {\bibfnamefont {C.}~\bibnamefont {Nayak}},\ }\bibfield  {title}
  {\bibinfo {title} {Prethermal strong zero modes and topological qubits},\
  }\href {https://doi.org/10.1103/PhysRevX.7.041062} {\bibfield  {journal}
  {\bibinfo  {journal} {Phys. Rev. X}\ }\textbf {\bibinfo {volume} {7}},\
  \bibinfo {pages} {041062} (\bibinfo {year} {2017}{\natexlab{a}})}\BibitemShut
  {NoStop}%
\bibitem [{\citenamefont {M{\"u}ller}\ and\ \citenamefont
  {Nersesyan}(2016)}]{Muller:2016}%
  \BibitemOpen
  \bibfield  {author} {\bibinfo {author} {\bibfnamefont {M.}~\bibnamefont
  {M{\"u}ller}}\ and\ \bibinfo {author} {\bibfnamefont {A.~A.}\ \bibnamefont
  {Nersesyan}},\ }\bibfield  {title} {\bibinfo {title} {{Classical impurities
  and boundary Majorana zero modes in quantum chains}},\ }\href
  {https://doi.org/10.1016/j.aop.2016.07.025} {\bibfield  {journal} {\bibinfo
  {journal} {Annals Phys.}\ }\textbf {\bibinfo {volume} {372}},\ \bibinfo
  {pages} {482} (\bibinfo {year} {2016})}\BibitemShut {NoStop}%
\bibitem [{\citenamefont {Kemp}\ \emph {et~al.}(2017)\citenamefont {Kemp},
  \citenamefont {Yao}, \citenamefont {Laumann},\ and\ \citenamefont
  {Fendley}}]{Kemp_2017}%
  \BibitemOpen
  \bibfield  {author} {\bibinfo {author} {\bibfnamefont {J.}~\bibnamefont
  {Kemp}}, \bibinfo {author} {\bibfnamefont {N.~Y.}\ \bibnamefont {Yao}},
  \bibinfo {author} {\bibfnamefont {C.~R.}\ \bibnamefont {Laumann}},\ and\
  \bibinfo {author} {\bibfnamefont {P.}~\bibnamefont {Fendley}},\ }\bibfield
  {title} {\bibinfo {title} {Long coherence times for edge spins},\ }\href
  {https://doi.org/10.1088/1742-5468/aa73f0} {\bibfield  {journal} {\bibinfo
  {journal} {Journal of Statistical Mechanics: Theory and Experiment}\ }\textbf
  {\bibinfo {volume} {2017}},\ \bibinfo {pages} {063105} (\bibinfo {year}
  {2017})}\BibitemShut {NoStop}%
\bibitem [{\citenamefont {Moran}\ \emph {et~al.}(2017)\citenamefont {Moran},
  \citenamefont {Pellegrino}, \citenamefont {Slingerland},\ and\ \citenamefont
  {Kells}}]{Moran17}%
  \BibitemOpen
  \bibfield  {author} {\bibinfo {author} {\bibfnamefont {N.}~\bibnamefont
  {Moran}}, \bibinfo {author} {\bibfnamefont {D.}~\bibnamefont {Pellegrino}},
  \bibinfo {author} {\bibfnamefont {J.~K.}\ \bibnamefont {Slingerland}},\ and\
  \bibinfo {author} {\bibfnamefont {G.}~\bibnamefont {Kells}},\ }\bibfield
  {title} {\bibinfo {title} {Parafermionic clock models and quantum
  resonance},\ }\href {https://doi.org/10.1103/PhysRevB.95.235127} {\bibfield
  {journal} {\bibinfo  {journal} {Phys. Rev. B}\ }\textbf {\bibinfo {volume}
  {95}},\ \bibinfo {pages} {235127} (\bibinfo {year} {2017})}\BibitemShut
  {NoStop}%
\bibitem [{\citenamefont {Vasiloiu}\ \emph {et~al.}(2018)\citenamefont
  {Vasiloiu}, \citenamefont {Carollo},\ and\ \citenamefont
  {Garrahan}}]{Vasiloiu18}%
  \BibitemOpen
  \bibfield  {author} {\bibinfo {author} {\bibfnamefont {L.~M.}\ \bibnamefont
  {Vasiloiu}}, \bibinfo {author} {\bibfnamefont {F.}~\bibnamefont {Carollo}},\
  and\ \bibinfo {author} {\bibfnamefont {J.~P.}\ \bibnamefont {Garrahan}},\
  }\bibfield  {title} {\bibinfo {title} {Enhancing correlation times for edge
  spins through dissipation},\ }\href
  {https://doi.org/10.1103/PhysRevB.98.094308} {\bibfield  {journal} {\bibinfo
  {journal} {Phys. Rev. B}\ }\textbf {\bibinfo {volume} {98}},\ \bibinfo
  {pages} {094308} (\bibinfo {year} {2018})}\BibitemShut {NoStop}%
\bibitem [{\citenamefont {Vasiloiu}\ \emph {et~al.}(2019)\citenamefont
  {Vasiloiu}, \citenamefont {Carollo}, \citenamefont {Marcuzzi},\ and\
  \citenamefont {Garrahan}}]{Vasiloiu19}%
  \BibitemOpen
  \bibfield  {author} {\bibinfo {author} {\bibfnamefont {L.~M.}\ \bibnamefont
  {Vasiloiu}}, \bibinfo {author} {\bibfnamefont {F.}~\bibnamefont {Carollo}},
  \bibinfo {author} {\bibfnamefont {M.}~\bibnamefont {Marcuzzi}},\ and\
  \bibinfo {author} {\bibfnamefont {J.~P.}\ \bibnamefont {Garrahan}},\
  }\bibfield  {title} {\bibinfo {title} {Strong zero modes in a class of
  generalized ising spin ladders with plaquette interactions},\ }\href
  {https://doi.org/10.1103/PhysRevB.100.024309} {\bibfield  {journal} {\bibinfo
   {journal} {Phys. Rev. B}\ }\textbf {\bibinfo {volume} {100}},\ \bibinfo
  {pages} {024309} (\bibinfo {year} {2019})}\BibitemShut {NoStop}%
\bibitem [{\citenamefont {Yates}\ \emph {et~al.}(2019)\citenamefont {Yates},
  \citenamefont {Essler},\ and\ \citenamefont {Mitra}}]{Yates19}%
  \BibitemOpen
  \bibfield  {author} {\bibinfo {author} {\bibfnamefont {D.~J.}\ \bibnamefont
  {Yates}}, \bibinfo {author} {\bibfnamefont {F.~H.~L.}\ \bibnamefont
  {Essler}},\ and\ \bibinfo {author} {\bibfnamefont {A.}~\bibnamefont
  {Mitra}},\ }\bibfield  {title} {\bibinfo {title} {Almost strong
  ($0,\ensuremath{\pi}$) edge modes in clean interacting one-dimensional
  floquet systems},\ }\href {https://doi.org/10.1103/PhysRevB.99.205419}
  {\bibfield  {journal} {\bibinfo  {journal} {Phys. Rev. B}\ }\textbf {\bibinfo
  {volume} {99}},\ \bibinfo {pages} {205419} (\bibinfo {year}
  {2019})}\BibitemShut {NoStop}%
\bibitem [{\citenamefont {Kemp}\ \emph {et~al.}(2020)\citenamefont {Kemp},
  \citenamefont {Yao},\ and\ \citenamefont {Laumann}}]{kemp2019symmetry}%
  \BibitemOpen
  \bibfield  {author} {\bibinfo {author} {\bibfnamefont {J.}~\bibnamefont
  {Kemp}}, \bibinfo {author} {\bibfnamefont {N.~Y.}\ \bibnamefont {Yao}},\ and\
  \bibinfo {author} {\bibfnamefont {C.~R.}\ \bibnamefont {Laumann}},\
  }\bibfield  {title} {\bibinfo {title} {Symmetry-enhanced boundary qubits at
  infinite temperature},\ }\href
  {https://doi.org/10.1103/PhysRevLett.125.200506} {\bibfield  {journal}
  {\bibinfo  {journal} {Phys. Rev. Lett.}\ }\textbf {\bibinfo {volume} {125}},\
  \bibinfo {pages} {200506} (\bibinfo {year} {2020})}\BibitemShut {NoStop}%
\bibitem [{\citenamefont {Rakovszky}\ \emph {et~al.}(2020)\citenamefont
  {Rakovszky}, \citenamefont {Sala}, \citenamefont {Verresen}, \citenamefont
  {Knap},\ and\ \citenamefont {Pollmann}}]{Rakovszky20}%
  \BibitemOpen
  \bibfield  {author} {\bibinfo {author} {\bibfnamefont {T.}~\bibnamefont
  {Rakovszky}}, \bibinfo {author} {\bibfnamefont {P.}~\bibnamefont {Sala}},
  \bibinfo {author} {\bibfnamefont {R.}~\bibnamefont {Verresen}}, \bibinfo
  {author} {\bibfnamefont {M.}~\bibnamefont {Knap}},\ and\ \bibinfo {author}
  {\bibfnamefont {F.}~\bibnamefont {Pollmann}},\ }\bibfield  {title} {\bibinfo
  {title} {Statistical localization: From strong fragmentation to strong edge
  modes},\ }\href {https://doi.org/10.1103/PhysRevB.101.125126} {\bibfield
  {journal} {\bibinfo  {journal} {Phys. Rev. B}\ }\textbf {\bibinfo {volume}
  {101}},\ \bibinfo {pages} {125126} (\bibinfo {year} {2020})}\BibitemShut
  {NoStop}%
\bibitem [{\citenamefont {Yeh}\ \emph {et~al.}(2023)\citenamefont {Yeh},
  \citenamefont {Rosch},\ and\ \citenamefont {Mitra}}]{yeh_decay_2023}%
  \BibitemOpen
  \bibfield  {author} {\bibinfo {author} {\bibfnamefont {H.-C.}\ \bibnamefont
  {Yeh}}, \bibinfo {author} {\bibfnamefont {A.}~\bibnamefont {Rosch}},\ and\
  \bibinfo {author} {\bibfnamefont {A.}~\bibnamefont {Mitra}},\ }\bibfield
  {title} {\bibinfo {title} {Decay rates of almost strong modes in {Floquet}
  spin chains beyond {Fermi}'s {Golden} {Rule}},\ }\href@noop {} {\  (\bibinfo
  {year} {2023})},\ \Eprint {https://arxiv.org/abs/2305.04980}
  {arXiv:2305.04980} \BibitemShut {NoStop}%
\bibitem [{top()}]{topologynote}%
  \BibitemOpen
  \href@noop {} {}\bibinfo {note} {The terminology is in reference to the
  well-known topological Majorana zero modes in the Kitaev
  chain~\cite{Kitaev_2001}, but we stress that many examples, such as the
  transverse-field Ising model (TFIM), do \emph{not} have corresponding
  topological edge modes in the ground state.}\BibitemShut {Stop}%
\bibitem [{\citenamefont {Abanin}\ \emph {et~al.}(2017)\citenamefont {Abanin},
  \citenamefont {De~Roeck}, \citenamefont {Ho},\ and\ \citenamefont
  {Huveneers}}]{Abanin17}%
  \BibitemOpen
  \bibfield  {author} {\bibinfo {author} {\bibfnamefont {D.}~\bibnamefont
  {Abanin}}, \bibinfo {author} {\bibfnamefont {W.}~\bibnamefont {De~Roeck}},
  \bibinfo {author} {\bibfnamefont {W.~W.}\ \bibnamefont {Ho}},\ and\ \bibinfo
  {author} {\bibfnamefont {F.}~\bibnamefont {Huveneers}},\ }\bibfield  {title}
  {\bibinfo {title} {A rigorous theory of many-body prethermalization for
  periodically driven and closed quantum systems},\ }\href@noop {} {\bibfield
  {journal} {\bibinfo  {journal} {Communications in Mathematical Physics}\
  }\textbf {\bibinfo {volume} {354}},\ \bibinfo {pages} {809} (\bibinfo {year}
  {2017})}\BibitemShut {NoStop}%
\bibitem [{\citenamefont {Else}\ \emph
  {et~al.}(2017{\natexlab{b}})\citenamefont {Else}, \citenamefont {Bauer},\
  and\ \citenamefont {Nayak}}]{PhysRevX.7.011026}%
  \BibitemOpen
  \bibfield  {author} {\bibinfo {author} {\bibfnamefont {D.~V.}\ \bibnamefont
  {Else}}, \bibinfo {author} {\bibfnamefont {B.}~\bibnamefont {Bauer}},\ and\
  \bibinfo {author} {\bibfnamefont {C.}~\bibnamefont {Nayak}},\ }\bibfield
  {title} {\bibinfo {title} {Prethermal phases of matter protected by
  time-translation symmetry},\ }\href
  {https://doi.org/10.1103/PhysRevX.7.011026} {\bibfield  {journal} {\bibinfo
  {journal} {Phys. Rev. X}\ }\textbf {\bibinfo {volume} {7}},\ \bibinfo {pages}
  {011026} (\bibinfo {year} {2017}{\natexlab{b}})}\BibitemShut {NoStop}%
\bibitem [{\citenamefont {Yates}\ \emph
  {et~al.}(2020{\natexlab{a}})\citenamefont {Yates}, \citenamefont {Abanov},\
  and\ \citenamefont {Mitra}}]{Yates20a}%
  \BibitemOpen
  \bibfield  {author} {\bibinfo {author} {\bibfnamefont {D.~J.}\ \bibnamefont
  {Yates}}, \bibinfo {author} {\bibfnamefont {A.~G.}\ \bibnamefont {Abanov}},\
  and\ \bibinfo {author} {\bibfnamefont {A.}~\bibnamefont {Mitra}},\ }\bibfield
   {title} {\bibinfo {title} {Lifetime of almost strong edge-mode operators in
  one-dimensional, interacting, symmetry protected topological phases},\ }\href
  {https://doi.org/10.1103/PhysRevLett.124.206803} {\bibfield  {journal}
  {\bibinfo  {journal} {Phys. Rev. Lett.}\ }\textbf {\bibinfo {volume} {124}},\
  \bibinfo {pages} {206803} (\bibinfo {year} {2020}{\natexlab{a}})}\BibitemShut
  {NoStop}%
\bibitem [{\citenamefont {Yates}\ \emph
  {et~al.}(2020{\natexlab{b}})\citenamefont {Yates}, \citenamefont {Abanov},\
  and\ \citenamefont {Mitra}}]{Yates20b}%
  \BibitemOpen
  \bibfield  {author} {\bibinfo {author} {\bibfnamefont {D.~J.}\ \bibnamefont
  {Yates}}, \bibinfo {author} {\bibfnamefont {A.~G.}\ \bibnamefont {Abanov}},\
  and\ \bibinfo {author} {\bibfnamefont {A.}~\bibnamefont {Mitra}},\ }\bibfield
   {title} {\bibinfo {title} {Dynamics of almost strong edge modes in spin
  chains away from integrability},\ }\href
  {https://doi.org/10.1103/PhysRevB.102.195419} {\bibfield  {journal} {\bibinfo
   {journal} {Phys. Rev. B}\ }\textbf {\bibinfo {volume} {102}},\ \bibinfo
  {pages} {195419} (\bibinfo {year} {2020}{\natexlab{b}})}\BibitemShut
  {NoStop}%
\bibitem [{\citenamefont {Alicea}\ \emph {et~al.}(2011)\citenamefont {Alicea},
  \citenamefont {Oreg}, \citenamefont {Refael}, \citenamefont {{von Oppen}},\
  and\ \citenamefont {Fisher}}]{alicea:2011}%
  \BibitemOpen
  \bibfield  {author} {\bibinfo {author} {\bibfnamefont {J.}~\bibnamefont
  {Alicea}}, \bibinfo {author} {\bibfnamefont {Y.}~\bibnamefont {Oreg}},
  \bibinfo {author} {\bibfnamefont {G.}~\bibnamefont {Refael}}, \bibinfo
  {author} {\bibfnamefont {F.}~\bibnamefont {{von Oppen}}},\ and\ \bibinfo
  {author} {\bibfnamefont {M.~P.~A.}\ \bibnamefont {Fisher}},\ }\bibfield
  {title} {\bibinfo {title} {Non-{{Abelian}} statistics and topological quantum
  information processing in {{1D}} wire networks},\ }\href
  {https://doi.org/10.1038/nphys1915} {\bibfield  {journal} {\bibinfo
  {journal} {Nature Physics}\ }\textbf {\bibinfo {volume} {7}},\ \bibinfo
  {pages} {412} (\bibinfo {year} {2011})}\BibitemShut {NoStop}%
\bibitem [{par()}]{parafermnote}%
  \BibitemOpen
  \href@noop {} {}\bibinfo {note} {Here we are assuming that $\mathcal{F}^2=1$;
  we note that in systems where $\mathcal{F}^n=1$ for $n>2$, these criteria
  will be appropriately modified (e.g. in parafermionic
  systems)~\cite{Fendley_2016,Alicea2016,Jermyn:2014}.}\BibitemShut {Stop}%
\bibitem [{\citenamefont {Jordan}\ and\ \citenamefont {Wigner}(1928)}]{JW1928}%
  \BibitemOpen
  \bibfield  {author} {\bibinfo {author} {\bibfnamefont {P.}~\bibnamefont
  {Jordan}}\ and\ \bibinfo {author} {\bibfnamefont {E.}~\bibnamefont
  {Wigner}},\ }\bibfield  {title} {\bibinfo {title} {{\"U}ber das paulische
  {\"a}quivalenzverbot},\ }\href {https://doi.org/10.1007/BF01331938}
  {\bibfield  {journal} {\bibinfo  {journal} {Zeitschrift f{\"u}r Physik}\
  }\textbf {\bibinfo {volume} {47}},\ \bibinfo {pages} {631} (\bibinfo {year}
  {1928})}\BibitemShut {NoStop}%
\bibitem [{\citenamefont {Fendley}(2014)}]{Fendley_2014}%
  \BibitemOpen
  \bibfield  {author} {\bibinfo {author} {\bibfnamefont {P.}~\bibnamefont
  {Fendley}},\ }\bibfield  {title} {\bibinfo {title} {Free parafermions},\
  }\href {https://doi.org/10.1088/1751-8113/47/7/075001} {\bibfield  {journal}
  {\bibinfo  {journal} {Journal of Physics A: Mathematical and Theoretical}\
  }\textbf {\bibinfo {volume} {47}},\ \bibinfo {pages} {075001} (\bibinfo
  {year} {2014})}\BibitemShut {NoStop}%
\bibitem [{sup()}]{suppinfo}%
  \BibitemOpen
  \href@noop {} {}\bibinfo {note} {See Supplementary Material for additional
  details.}\BibitemShut {Stop}%
\bibitem [{\citenamefont {Schultz}\ \emph {et~al.}(1964)\citenamefont
  {Schultz}, \citenamefont {Mattis},\ and\ \citenamefont {Lieb}}]{SML}%
  \BibitemOpen
  \bibfield  {author} {\bibinfo {author} {\bibfnamefont {T.~D.}\ \bibnamefont
  {Schultz}}, \bibinfo {author} {\bibfnamefont {D.~C.}\ \bibnamefont
  {Mattis}},\ and\ \bibinfo {author} {\bibfnamefont {E.~H.}\ \bibnamefont
  {Lieb}},\ }\bibfield  {title} {\bibinfo {title} {Two-dimensional ising model
  as a soluble problem of many fermions},\ }\href
  {https://doi.org/10.1103/RevModPhys.36.856} {\bibfield  {journal} {\bibinfo
  {journal} {Rev. Mod. Phys.}\ }\textbf {\bibinfo {volume} {36}},\ \bibinfo
  {pages} {856} (\bibinfo {year} {1964})}\BibitemShut {NoStop}%
\bibitem [{\citenamefont {Bravyi}\ and\ \citenamefont
  {K{\"o}nig}(2012)}]{bravyi:2012}%
  \BibitemOpen
  \bibfield  {author} {\bibinfo {author} {\bibfnamefont {S.}~\bibnamefont
  {Bravyi}}\ and\ \bibinfo {author} {\bibfnamefont {R.}~\bibnamefont
  {K{\"o}nig}},\ }\bibfield  {title} {\bibinfo {title} {Disorder-{{Assisted
  Error Correction}} in {{Majorana Chains}}},\ }\href
  {https://doi.org/10.1007/s00220-012-1606-9} {\bibfield  {journal} {\bibinfo
  {journal} {Commun. Math. Phys.}\ }\textbf {\bibinfo {volume} {316}},\
  \bibinfo {pages} {641} (\bibinfo {year} {2012})}\BibitemShut {NoStop}%
\bibitem [{osc()}]{oscillatenote}%
  \BibitemOpen
  \href@noop {} {}\bibinfo {note} {In fact, due to the system's free-fermion
  nature, the autocorrelator never decays, but rather starts to coherently
  oscillate.}\BibitemShut {Stop}%
\bibitem [{\citenamefont {Petersen}(2015)}]{petersen:2015}%
  \BibitemOpen
  \bibfield  {author} {\bibinfo {author} {\bibfnamefont {T.~K.}\ \bibnamefont
  {Petersen}},\ }\href {https://doi.org/10.1007/978-1-4939-3091-3} {\emph
  {\bibinfo {title} {Eulerian {{Numbers}}}}},\ Birkh\"auser {{Advanced Texts}}
  {{Basler Lehrb\"ucher}}\ (\bibinfo  {publisher} {{Birkh\"auser Basel}},\
  \bibinfo {year} {2015})\BibitemShut {NoStop}%
\bibitem [{\citenamefont {Meyer}(2020)}]{meyer:2020}%
  \BibitemOpen
  \bibfield  {author} {\bibinfo {author} {\bibfnamefont {G.}~\bibnamefont
  {Meyer}},\ }\href {https://doi.org/10.5281/zenodo.3606826} {\bibinfo {title}
  {{{GregDMeyer}}/dynamite v0.1.0}},\ \bibinfo {howpublished} {Zenodo}
  (\bibinfo {year} {2020})\BibitemShut {NoStop}%
\bibitem [{\citenamefont {Grabowski}\ and\ \citenamefont
  {Mathieu}(1994)}]{grabowski:1994}%
  \BibitemOpen
  \bibfield  {author} {\bibinfo {author} {\bibfnamefont {M.~P.}\ \bibnamefont
  {Grabowski}}\ and\ \bibinfo {author} {\bibfnamefont {P.}~\bibnamefont
  {Mathieu}},\ }\bibfield  {title} {\bibinfo {title} {Quantum integrals of
  motion for the heisenberg spin chain},\ }\href
  {https://doi.org/10.1142/S0217732394002057} {\bibfield  {journal} {\bibinfo
  {journal} {Mod. Phys. Lett. A}\ }\textbf {\bibinfo {volume} {09}},\ \bibinfo
  {pages} {2197} (\bibinfo {year} {1994})}\BibitemShut {NoStop}%
\bibitem [{\citenamefont {Nozawa}\ and\ \citenamefont
  {Fukai}(2020)}]{nozawa:2020}%
  \BibitemOpen
  \bibfield  {author} {\bibinfo {author} {\bibfnamefont {Y.}~\bibnamefont
  {Nozawa}}\ and\ \bibinfo {author} {\bibfnamefont {K.}~\bibnamefont {Fukai}},\
  }\bibfield  {title} {\bibinfo {title} {Explicit {{Construction}} of {{Local
  Conserved Quantities}} in the {{XYZ}} {{Spin}}-$1/2$ {{Chain}}},\ }\href
  {https://doi.org/10.1103/PhysRevLett.125.090602} {\bibfield  {journal}
  {\bibinfo  {journal} {Phys. Rev. Lett.}\ }\textbf {\bibinfo {volume} {125}},\
  \bibinfo {pages} {090602} (\bibinfo {year} {2020})}\BibitemShut {NoStop}%
\bibitem [{\citenamefont {Nienhuis}\ and\ \citenamefont
  {Huijgen}(2021)}]{nienhuis:2021}%
  \BibitemOpen
  \bibfield  {author} {\bibinfo {author} {\bibfnamefont {B.}~\bibnamefont
  {Nienhuis}}\ and\ \bibinfo {author} {\bibfnamefont {O.}~\bibnamefont
  {Huijgen}},\ }\bibfield  {title} {\bibinfo {title} {The local conserved
  quantities of the closed {{XXZ}} chain},\ }\href
  {https://doi.org/10.1088/1751-8121/ac0961} {\bibfield  {journal} {\bibinfo
  {journal} {J. Phys. A: Math. Theor.}\ }\textbf {\bibinfo {volume} {54}},\
  \bibinfo {pages} {304001} (\bibinfo {year} {2021})}\BibitemShut {NoStop}%
\end{thebibliography}%


%


\begin{thebibliography}{5}%
\makeatletter
\providecommand \@ifxundefined [1]{%
 \@ifx{#1\undefined}
}%
\providecommand \@ifnum [1]{%
 \ifnum #1\expandafter \@firstoftwo
 \else \expandafter \@secondoftwo
 \fi
}%
\providecommand \@ifx [1]{%
 \ifx #1\expandafter \@firstoftwo
 \else \expandafter \@secondoftwo
 \fi
}%
\providecommand \natexlab [1]{#1}%
\providecommand \enquote  [1]{``#1''}%
\providecommand \bibnamefont  [1]{#1}%
\providecommand \bibfnamefont [1]{#1}%
\providecommand \citenamefont [1]{#1}%
\providecommand \href@noop [0]{\@secondoftwo}%
\providecommand \href [0]{\begingroup \@sanitize@url \@href}%
\providecommand \@href[1]{\@@startlink{#1}\@@href}%
\providecommand \@@href[1]{\endgroup#1\@@endlink}%
\providecommand \@sanitize@url [0]{\catcode `\\12\catcode `\$12\catcode
  `\&12\catcode `\#12\catcode `\^12\catcode `\_12\catcode `\%12\relax}%
\providecommand \@@startlink[1]{}%
\providecommand \@@endlink[0]{}%
\providecommand \url  [0]{\begingroup\@sanitize@url \@url }%
\providecommand \@url [1]{\endgroup\@href {#1}{\urlprefix }}%
\providecommand \urlprefix  [0]{URL }%
\providecommand \Eprint [0]{\href }%
\providecommand \doibase [0]{http://dx.doi.org/}%
\providecommand \selectlanguage [0]{\@gobble}%
\providecommand \bibinfo  [0]{\@secondoftwo}%
\providecommand \bibfield  [0]{\@secondoftwo}%
\providecommand \translation [1]{[#1]}%
\providecommand \BibitemOpen [0]{}%
\providecommand \bibitemStop [0]{}%
\providecommand \bibitemNoStop [0]{.\EOS\space}%
\providecommand \EOS [0]{\spacefactor3000\relax}%
\providecommand \BibitemShut  [1]{\csname bibitem#1\endcsname}%
\let\auto@bib@innerbib\@empty
\bibitem [{\citenamefont {Else}\ \emph {et~al.}(2017)\citenamefont {Else},
  \citenamefont {Fendley}, \citenamefont {Kemp},\ and\ \citenamefont
  {Nayak}}]{SElse17}%
  \BibitemOpen
  \bibfield  {author} {\bibinfo {author} {\bibfnamefont {D.~V.}\ \bibnamefont
  {Else}}, \bibinfo {author} {\bibfnamefont {P.}~\bibnamefont {Fendley}},
  \bibinfo {author} {\bibfnamefont {J.}~\bibnamefont {Kemp}}, \ and\ \bibinfo
  {author} {\bibfnamefont {C.}~\bibnamefont {Nayak}},\ }\href {\doibase
  10.1103/PhysRevX.7.041062} {\bibfield  {journal} {\bibinfo  {journal} {Phys.
  Rev. X}\ }\textbf {\bibinfo {volume} {7}},\ \bibinfo {pages} {041062}
  (\bibinfo {year} {2017})}\BibitemShut {NoStop}%
\bibitem [{\citenamefont {Petersen}(2015)}]{Spetersen:2015}%
  \BibitemOpen
  \bibfield  {author} {\bibinfo {author} {\bibfnamefont {T.~K.}\ \bibnamefont
  {Petersen}},\ }\href {\doibase 10.1007/978-1-4939-3091-3} {\emph {\bibinfo
  {title} {Eulerian {{Numbers}}}}},\ Birkh\"auser {{Advanced Texts}} {{Basler
  Lehrb\"ucher}}\ (\bibinfo  {publisher} {{Birkh\"auser Basel}},\ \bibinfo
  {year} {2015})\BibitemShut {NoStop}%
\bibitem [{\citenamefont {Lando}\ and\ \citenamefont
  {Zvonkin}(1993)}]{Slando_plane_1993}%
  \BibitemOpen
  \bibfield  {author} {\bibinfo {author} {\bibfnamefont {S.~K.}\ \bibnamefont
  {Lando}}\ and\ \bibinfo {author} {\bibfnamefont {A.~K.}\ \bibnamefont
  {Zvonkin}},\ }\href {\doibase 10.1016/0304-3975(93)90316-L} {\bibfield
  {journal} {\bibinfo  {journal} {Theoretical Computer Science}\ }\textbf
  {\bibinfo {volume} {117}},\ \bibinfo {pages} {227} (\bibinfo {year}
  {1993})}\BibitemShut {NoStop}%
\bibitem [{\citenamefont {Ackerhalt}\ and\ \citenamefont {Rza\ifmmode
  \mbox{\c{}}\else \c{}\fi{}\ifmmode~\dot{z}\else
  \.{z}\fi{}ewski}(1975)}]{SPhysRevA.12.2549}%
  \BibitemOpen
  \bibfield  {author} {\bibinfo {author} {\bibfnamefont {J.~R.}\ \bibnamefont
  {Ackerhalt}}\ and\ \bibinfo {author} {\bibfnamefont {K.}~\bibnamefont
  {Rza\ifmmode \mbox{\c{}}\else \c{}\fi{}\ifmmode~\dot{z}\else
  \.{z}\fi{}ewski}},\ }\href {\doibase 10.1103/PhysRevA.12.2549} {\bibfield
  {journal} {\bibinfo  {journal} {Phys. Rev. A}\ }\textbf {\bibinfo {volume}
  {12}},\ \bibinfo {pages} {2549} (\bibinfo {year} {1975})}\BibitemShut
  {NoStop}%
\bibitem [{\citenamefont {Kemp}\ \emph {et~al.}(2017)\citenamefont {Kemp},
  \citenamefont {Yao}, \citenamefont {Laumann},\ and\ \citenamefont
  {Fendley}}]{SKemp_2017}%
  \BibitemOpen
  \bibfield  {author} {\bibinfo {author} {\bibfnamefont {J.}~\bibnamefont
  {Kemp}}, \bibinfo {author} {\bibfnamefont {N.~Y.}\ \bibnamefont {Yao}},
  \bibinfo {author} {\bibfnamefont {C.~R.}\ \bibnamefont {Laumann}}, \ and\
  \bibinfo {author} {\bibfnamefont {P.}~\bibnamefont {Fendley}},\ }\href
  {\doibase 10.1088/1742-5468/aa73f0} {\bibfield  {journal} {\bibinfo
  {journal} {Journal of Statistical Mechanics: Theory and Experiment}\ }\textbf
  {\bibinfo {volume} {2017}},\ \bibinfo {pages} {063105} (\bibinfo {year}
  {2017})}\BibitemShut {NoStop}%
\end{thebibliography}

\onecolumngrid
\clearpage

\widetext

\setcounter{equation}{0}
\setcounter{figure}{0}
\setcounter{table}{0}
\setcounter{page}{1}
\makeatletter
\renewcommand{\theequation}{S\arabic{equation}}
\renewcommand{\thefigure}{S\arabic{figure}}
\renewcommand{\bibnumfmt}[1]{[S#1]}
\renewcommand{\citenumfont}[1]{S#1}


\newcommand{\sgn}{\operatorname{sgn}}
\newcommand{\eff}{\mathrm{eff}}
\newcommand{\local}{{\mathrm{local}}}
\newcommand{\static}{{\mathrm{static}}}
\newcommand{\drive}{{\mathrm{drive}}}
\newcommand{\expct}[1]{\langle #1 \rangle}
\def\X{\sigma^x}
\def\Y{\sigma^y}
\def\Z{\sigma^z}
\def \ab{\alpha^\prime}
\def \abb{\alpha^{\prime\prime}}
\def \bb{\beta^\prime}
\def \bbb{\beta^{\prime\prime}}
\def \nop{-1^{\prime}}

\begin{center}
\textbf{\large Supplementary Information:\\\vspace{1mm} Boundary Strong Zero Modes}

\vspace{10pt}
\thispagestyle{plain}
Christopher T. Olund,\textsuperscript{1} Norman Y. Yao,\textsuperscript{1, 2} and Jack Kemp\textsuperscript{2}

\textsuperscript{1}\textit{\small Department of Physics, University of California at Berkeley, Berkeley, CA 94720, USA}

\textsuperscript{2}\textit{\small Department of Physics, Harvard University, MA 02138, USA}
\end{center}
\author{Christopher T. Olund}
\affiliation{}
\author{Norman Y. Yao}
\affiliation{Department of Physics, University of California at Berkeley, Berkeley, CA 94720, USA}
\affiliation{Department of Physics, Harvard University, MA 02138, USA}
\author{Jack Kemp}
\affiliation{Department of Physics, Harvard University, MA 02138, USA}

\section{Choice of Jordan--Wigner String}
In the main text, we choose a Jordan--Wignerization starting from the left edge of the system. This means that the operator in the spin boundary Ising model corresponding to $a_0$ in the fermionic boundary Ising model is not simply $\sigma_0^z$, but instead the highly nonlocal operator $\left(\prod_{k=-\infty}^{-1}\sigma_k^x\right)\sigma_0^z$. Starting from the fermionic Hamiltonian, a natural choice for a different set of JW transformations is one that maintains an exact correspondence between $a_0$ and $\sigma^z_0$. In order to do this, we define a new string operator that starts at the center of the chain at site $0$ and then wraps around from $+\infty$ to $-\infty$:
\begin{align}
    a_j= & \begin{cases}
    \left(\prod_{k=-\infty}^{j-1}\sigma_k^x\right)\left(\prod_{k=0}^{\infty}\sigma_k^x\right)\sigma_j^z & j < 0\\
    \left(\prod_{k=0}^{j-1}\sigma_k^x\right)\sigma_j^z & j\ge 0
    \end{cases}\\
    b_j= & \begin{cases}
    i\left(\prod_{k=-\infty}^{j}\sigma_k^x\right)\left(\prod_{k=0}^{\infty}\sigma_k^x\right)\sigma_j^z & j < 0\\
    i\left(\prod_{k=0}^{j}\sigma_k^x\right)\sigma_j^z & j\ge 0
    \end{cases}.
\end{align}

Under such a JW transformation, $a_0$ trivially transforms into $\sigma^z_0$, allowing us to directly compare the strong zero modes living on the boundary. However, the Hamiltonian $H_\text{BI}$ also changes:
\begin{equation}
    H^{\prime}_\text{BI}=-J_1\sum_{j=-\infty}^{-2}\sigma_j^z\sigma_{j+1}^z
    -J_2\sum_{j=0}^{\infty}\sigma_j^z\sigma_{j+1}^z
    -h_1\sum_{j=-\infty}^{-1}\sigma_j^x-h_2\sum_{j=0}^\infty\sigma_j^x
    -J_1\left(\prod_{k=-\infty}^{\infty}\sigma_k^x\right)\sigma_{-1}^z\sigma_0^z.
\end{equation}

The JW string attached to $\sigma^z_{-1}$ is no longer cancelled by that attached to $\sigma^z_{0}$, and thus the coupling between the two chains picks up a nonlocal factor of a global spin flip about the $\hat{x}$ axis, the symmetry operator $\mathcal{F}$. This factor drives a significant difference in physics between the spin and fermionic SZMs, as we have seen in the main text.

\section{Exact Solution for the spin boundary SZM}
In this Appendix, we shall prove the existence of an exact SZM starting at the boundary spin $\sigma^z_0$ between two transverse field Ising chains with different couplings. In particular we focus on the same Hamiltonian as in the main text but with the left-hand chain a finite $N$ sites long:

\begin{equation}
    H_\textrm{BI}=-J_1\sum_{j=-N}^{-1}\sigma_j^z\sigma_{j+1}^z
    -J_2\sum_{j=0}^{\infty}\sigma_j^z\sigma_{j+1}^z
    -h_1\sum_{j=-\infty}^{-1}\sigma_j^x-h_2\sum_{j=0}^\infty\sigma_j^x.
\end{equation}
We will solve the case $N=1$ exactly, and prove that in the limit as $N \to \infty$ the SZM is exponentially localized under a set of constraints~\eqref{eq:condition} for the couplings that we derive. Additionally, we provide evidence that under these restrictions the its operator norm (hereafter `normalization'), converges.

We will often consider the same Hamiltonian written in terms of $a_i, b_i$, the usual Majorana fermion operators defined with a Jordan--Wigner string starting from the left edge of the chain, see Eq.~2 in the main text:
\begin{equation}
    H_\text{FBI}=-iJ_1\sum_{j=-N}^{-1}b_ja_{j+1}
    -iJ_2\sum_{j=0}^{\infty}b_ja_{j+1}\\
    -ih_1\sum_{j=-\infty}^{-1}a_jb_j-ih_2\sum_{j=0}^\infty a_jb_j.
\end{equation}

\subsection{Ansatz}
We will use the following ansatz for the SZM:
\begin{equation}
    \Psi_{N}=i^N\mathcal{N}\sum_{\mathclap{\substack{i_0<\dots<i_N\\j_0<\dots<j_{N-1}}}}A_{i_0\dots i_N} B_{j_0 \dots j_{N-1}}a_{i_0}  b_{j_0} a_{i_1} b_{j_1}\dots a_{i_N} ,
    \label{eq:ansatz}
\end{equation}
for real scalar coefficients $A, B$ and normalization $\mathcal{N}$. Additionally, to ensure the normalization $\mathcal{N}$ measures the overlap with the boundary spin $\sigma^z_0 = i^N (\prod^{-1}_{j=-N} a_j b_j) a_0$, we fix the coefficent of this term in the sum to one: $A_{-N,-N+1,\dots,0}=B_{-N,-N+1,\dots, -1}=1.$

The use of this ansatz enforces two major assumptions: firstly, every term in the SZM expansion has the same number of $a$ and $b$ Majorana fermion operators; secondly, the magnitude of each term in the expansion can be factorized into two contributions which only depend on which $a$ or $b$ operators are present independently. As observed in the main text, the \emph{total} number of Majorana operators remaining constant in each term of $\Psi_N$ immediately follows from the quadratic, free-fermion nature of $H_\text{BI}.$ The further separate conservation of $a$ and $b$ operator number is natural if we insist that, as well as commuting with the Hamiltonian, the SZM must be related to the boundary spin by a unitary transform: $\Psi_N=U^\dag \sigma^z_0 U$. This ensures that the SZM squares to the identity, $\Psi_N^2 = 1$. It is also consistent with the understanding that the SZM is a consequence of the emergent U(1) symmetry revealed by a local unitary transformation of the Hamiltonian to a new Hamiltonian under which the boundary spin is conserved exactly\cite{Else17}.

In order to explain the separate conservation of $a$ and $b$ operator number, let us first assume the converse, so that $\Psi_N$ consists of the sums of arbitrary strings of $a$ and $b$ operators total length $2N+1$. Let the number of $a$ operators in a given string $s$ be $N_a(s)$, and let the commutator $C = [H_\text{BI}, \Psi_N].$ For $\Psi_N$ to commute with the Hamiltonian every term in $C$ must individually vanish. Notice that one commutation with $H_{BI}$ changes the number of $a$ operators in a string by $\pm 1$; thus, each operator string $s$ with $N_a(s) = n$ in $C$ originated from an operator string $s^\prime$ with $N_a(s^\prime) = n \pm 1$ in $\Psi_N$. This means that enforcing commutation with the Hamiltonian only relates the coefficients of operator strings in $\Psi_N$ with the same parity of $N_a(s)$. We can thus neglect any operator string in $\Psi_N$ with opposite $N_a$-parity to the zeroth order term $\sigma^z_0$. 

Having established this condition on operator strings in $\Psi_N$, let us turn to the form of the unitary transform $U$. Without loss of generality, we may write as $U=e^{G}$ for anti-Hermitian $G$. Every possible operator string with the same total Majorana fermion operator number as $\sigma^z_0$ can be generated by a unitary transform defined by \begin{equation} G = \sum^\infty_{i, j = -N}\left(\alpha_{ij} a_i a_j + \beta_{ij} b_i b_j + \gamma_{ij} a_i b_j\right),\label{eq:unitary}\end{equation} for arbitrary constants $\alpha_{ij}$, $\beta_{ij}$ and $\gamma_{ij}$. However, the $a_i b_j$ terms in $G$ will result in operator strings in $\Psi_N$ of arbitrary $N_a$-parity; thus, from the discussion above, we must set $\gamma_{ij}=0$. Possible additional terms in $G$ greater than quadratic in Majorana fermion operators, which could feasibly conserve $N_a$-parity by changing $N_a(\sigma^z_0)$ by an even number, such as $a_i a_j b_k b_l$, will generically not conserve total Majorana fermion operator number. The remaining terms in $G$ conserve the number of $a$ and $b$ operators exactly. Furthermore, notice that if $\gamma_{ij}=0$, $U$ splits into two mutually commuting parts: $U=U_a U_b = e^{G_a} e^{G_b}$, where $G_{a} = \sum \alpha_{ij} a_i a_j$ and  $G_{b} = \sum \beta_{ij} b_i b_j$. This justifies the factorization of the magnitude of each term in our ansatz for $\Psi_N$ into two contributions $A$ and $B$ which only depend on which $a$ or $b$ operators are present respectively.

\subsection{Counting terms involving a fixed number of hops}
Having defined the ansatz, we now wish to consider how many terms there are of a given order when we send $N$ to infinity. 
The order in perturbation theory each term in this ansatz appears is given by the minimum number of `hops' each Majorana fermion operator $a_i$ and $b_j$ must make on their respective sublattices to reach that term from their initial positions at zeroth order, $S a_0$, where $S$ is the product of all $a_i$s and $b_i$s for $i<0$. For example, in the ferromagnetic regime of both chains, a fermion-operator product that can be reached from $Sa_0$ in no fewer than $n$ hops will appear at leading order  $\propto (\prod_\text{j}h(j)J(j))/(J_1^2-J_2^2)^n$, where $h(j)$ and $J(j)$ are the $h$ and $J$ associated with hop $j$ of $n$. One can calculate the coefficients using the appropriate $G$, which we find to be
\begin{equation}
    G\approx \sum_{n=0}^\infty\frac{1}{2(J_1^2-J_2^2)^n}
    \sum_{m=0}^{n-1} (-1)^{m+1}\binom{n-1}{m}
    ((J_1h_1)^m
    (J_2h_2)^{n-m}a_{-m}a_{n-m}+
    (h_2/h_1)(J_1h_1)^{n-m}
    (J_2h_2)^m b_{m-n}b_{m})
\end{equation}
to term-wise leading order in $(J_1^2-J_2^2)^{-1}$.

In particular, we can map terms of order $n$ with $m$ `$a$' hops and $n-m$ `$b$' hops to simultaneous partitions of the integers $m$ and $n-m$. To see this, note that any given `$a$' or `$b$' can't have made any more hops than a fermion of the same type to its right; therefore, we can associate any partition with a collection of hops of one type of fermion uniquely, assigning the largest integer to the rightmost fermion of that flavor, the second largest integer to the fermion immediately to its left, and so on.
The number of partitions $p(n)$ of an integer $n$ is well known to be bounded by $(1/4\sqrt{3}n)\exp(\pi\sqrt{2n/3})$. We can then bound the number of $n$ hop terms $q(n)$ for even $n\ge 8$ by
\begin{equation}
    q(n)=\sum_{m=0}^n p(m)p(n-m) < n p\left(\frac{n}{2}\right)^2 < \frac{1}{12n}\exp\left(\pi\sqrt{\frac{4n}{3}}\right)\label{eq:countingterms}
\end{equation}
which is subexponential in $n$. This inequality in fact also holds for odd $n\ge 9$ if we replace $p(n/2)^2$ in the middle expression with $p((n-1)/2)p((n+1)/2)$. If the coefficients of the terms themselves are also at most exponential in $n$ (i.e. the SZM `exponentially localized to the boundary'), the norm of the operator formed by just keeping the leading order coefficients of terms in the SZM can be bounded by a geometric series, and will therefore be finite within some radius of convergence.

In the following discussion, we move beyond perturbation theory. For the toy model case of a single boundary spin, we calculate the SZM and its normalization exactly to prove convergence. For the general case, we are not able to calculate the SZM exactly, but instead provide sharp bounds on when it can be exponentially localized to the boundary (without any assumptions of keeping only leading-order terms in perturbation theory). The subexponential number of terms displayed here will then allow us to determine that under these same bounds the SZM normalization converges.

\subsection{Single boundary spin}
We turn to dealing exactly with the case $N=1$, that is, when there is a single spin past the boundary. In this case, our ansatz becomes
\begin{equation}\Psi_3 = i \mathcal{N} \sum^\infty_{l=-1} \sum^
\infty_{j=-1}\sum^\infty_{k=j+1}A_{j,k} B_l a_j a_k b_l,\label{eq:ansatz3}\end{equation} where we set $A_{-1,0}=B_{-1}=1.$

\begin{figure} [htbp]
\includegraphics[width=0.6\linewidth]{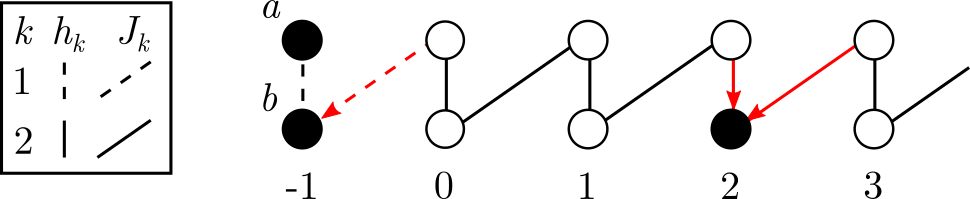}
\caption{The links of the chain depict the Hamiltonian couplings, while the filled circles denote an operator string, in this case $a_{-1} b_{-1} b_2$. The red arrows show how this string may be obtained from strings in $\Psi_3$ by the commutation $[H,\Psi_3]$. }
 \label{fig:3Ahopping}
\end{figure}

We want to find the appropriate $A_{j,k}, B_l$ such that each term in the commutator $C = [H_\text{BI}, \Psi_3]$ vanishes. Given the form of the ansatz and the Hamiltonian, operator strings in $C$ necessarily have the form $a_m b_n b_p$ or $a_m a_n a_p$. In order to visualize the contributions to each term in $C$, it is helpful to represent operator strings as diagrams as in Fig.~\ref{fig:3Ahopping}. Splitting each lattice site in two, the presence of Majorana fermion operators $a$ and $b$ on a lattice site in a string are represented by filled circles on the upper or lower chain respectively. The action of commuting the string by $H_\textrm{BI}$ amounts to `hopping' filled circles in this diagram to neighboring sites, with a hard-core constraint.

In particular, the diagram in Fig.~\ref{fig:3Ahopping} shows the contributions to the operator strings in $C$ of the form $a_{-1} b_{-1} b_m$ for $m \geq 0$. Here two of the Majorana operators remain in their original positions in $\sigma^z_0$ and one moves away. Demanding these terms vanish thus leads to the conditions:
\begin{align}J_2 A_{-1,m+1}&-h_2 A_{-1,m}-J_1 B_m=0 \nonumber  \\
A_{-1,m+1}&=\frac{h_2}{J_2} A_{-1,m}+\frac{J_1} {J_2}B_m \qquad m\geq 0 \label{eq:arec3} \end{align}
Recursively substituting the left hand side of equation \eqref{eq:arec3} into the right hand side, we may eliminate the $A_{1,j}$ entirely from the right hand side to find:
\begin{equation}
    A_{-1,m+1} = \sum^m_{j=0} \frac{h_2^{m-j}}{J_2^{m-j+1}}J_1 B_j +\left(\frac{h_2}{J_2}\right)^{m+1}  \qquad m \geq 0,
    \label{eq:am1m}
\end{equation}
where we have used $A_{-1,0}=1$.
We find $A_{0, m}$ from similar constraints on strings in $C$ of the form $a_{0} b_{-1} b_m$:
\begin{align}
    &J_2 A_{0,m+1}-h_2 A_{0,m}-h_1 B_m=0 \nonumber \qquad m\geq 1  \\
    &J_2 A_{0,1}-h_1 B_0=0 \label{eq:a01} \\
   &\implies A_{0, m+1} = \sum^m_{j=0} \frac{h_2^{m-j}}{J_2^{m-j+1}}h_1 B_j \qquad m \geq 0.
   \label{eq:a0m}
\end{align}

\begin{figure} [tbp]
\includegraphics[width=0.6\linewidth]{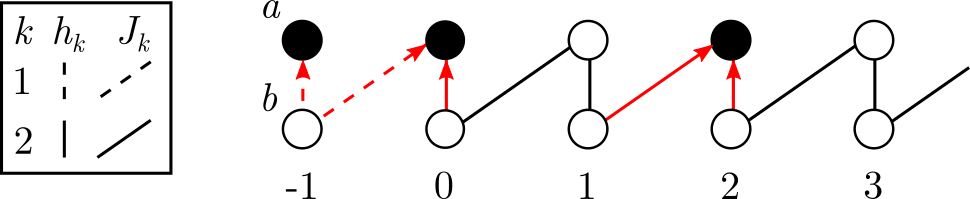}
\caption{As in Fig. 1, the filled circles denote an operator string, in this case $a_{-1} a_{-0} a_2$. The red arrows show how this string may be obtained from strings in $\Psi_3$ by the commutation $[H,\Psi_3]$. }
 \label{fig:bhopping3}
\end{figure}

Continuing to focus on strings in $C$ with only one Majorana fermion operator different from $\sigma^z_0$, let us consider those of form $a_{-1} a_0 a_m$ for $m>0$ (Fig~\ref{fig:bhopping3}). These impose the conditions:
\begin{equation}
J_2 B_{m-1}-h_2 B_{m}-h_1 A_{0,m}-J_1 A_{-1, m}+h_2 B_{0} A_{-1,m}=0 \qquad m\geq 1 \label{eq:baterm}
\end{equation}

Rewriting $A_{0,m}$ and $A_{1,m}$ using equations~\eqref{eq:am1m} and \eqref{eq:a0m} we find:
\begin{align}
h_2 B_{m} = J_2 B_{m-1}+\sum^{m-1}_{j=0} \left[(h_2 B_0-J_1) \frac{h_2^{m-j-1}}{J_2^{m-j}}J_1 - \frac{h_2^{m-j-1}}{J_2^{m-j}}h_1^2 \right]B_j+(h_2 B_0-J_1) \left(\frac{h_2}{J_2}\right)^m. \label{eq:b3complex}\end{align}
This is a recurrence relation for $B_m$ in terms of $B_j$ with $j<m$, suggesting we may be able to solve for arbitrary $B_m$ in terms of some initial set. Considerable simplification is possible if we first extract the last, $(m-1)$-th term from the sum on the right hand side:
\begin{align*}
    h_2  B_{m} &= \frac{1}{J_2}(J_2^2+h_2 B_0 J_1-h_1^2) B_{m-1}+\frac{h_2}{J_2}\left(\sum^{m-2}_{j=0} \left[(h_2 B_0-J_1) \frac{h_2^{m-j-2}}{J_2^{m-j-1}}J_1 - \frac{h_2^{m-j-2}}{J_2^{m-j-1}}h_1^2 \right]B_j +(h_2 B_0-J_1) \left(\frac{h_2}{J_2}\right)^{m-1} \right) 
\end{align*}
We can now replace the term in the large brackets using equation~\eqref{eq:b3complex} but with $m \to (m-1)$:
\begin{align}
    h_2 B_m &= \frac{1}{J_2}(J_2^2+h_2 B_0 J_1-h_1^2) B_{m-1}+\frac{h_2}{J_2}\left(h_2 B_{m-1}-J_2 B_{m-2}\right) \nonumber \\
    B_m & = \alpha B_{m-1} - B_{m-2}, \label{eq:threefermionrec}
\end{align}
for $\alpha = \frac{1}{h_2 J_2} (J_2^2-J_1^2+h_2^2-h_1^2+h_2 J_1 B_0)$. Equation \eqref{eq:threefermionrec} is a recurrence relation with constant coefficents and two unknown initial conditions, $B_0$ and $B_1$. However, $B_1$ can be directly calculated in terms of $B_0$ from equation~\eqref{eq:b3complex} with $m=1$ as $B_1 = \alpha B_0 - J_1/J_2.$
Thus, using basic combinatorics, we can now immediately solve for arbitrary $B_m$ in terms of $B_0$.

The most illuminating form of the solution for our purposes will be the generating function for the $B_m$, which is simply
\begin{equation}
    G(x) = \sum^\infty_{j=0} B_j x^j = \frac{B_0-(J_1/J_2)x}{1-\alpha x + x^2}.\label{eq:genfunc}
\end{equation}
Notice that the polynomial on the denominator is palindromic: it is self-reciprocal. Thus its roots can be written as $\lambda $ and $\lambda^{-1}$. Let us assume for the moment that both roots are real, and choose $\lambda \geq 1$, so that $\lambda^{-1} \leq 1$. Factoring the denominator as $(1-\lambda x)(1-\lambda^{-1} x)$ and expanding $G(x)$ as partial fractions, we find:
\begin{align*}G(x) &= \frac{c_1}{1-\lambda x} +\frac{c_2}{1-\lambda^{-1} x}\\
&= c_1(1+\lambda x + \lambda^2 x^2+\dots) + c_2(1+\lambda^{-1} x + \lambda^{-2} x^2+\dots),\end{align*}
for constants $c_1, c_2$. Because $\lambda > 1$, $G(x)$ diverges, which means that $B_m$ becomes larger and larger with $m$, and the SZM is delocalized. In order to avoid this fate, we must choose $B_0$ such that the numerator polynomial in equation~\eqref{eq:genfunc} \emph{cancels} the factor of $(1-\lambda x)$ in the denominator:
\begin{equation}\frac{B_0-(J_1/J_2)x}{(1-\lambda x)}=B_0 \qquad \forall x,\label{eq:cancellation}\end{equation} where the constant on the right hand side must be $B_0$ for $G(0)= B_0$. Notice also $\lambda$ depends on $B_0$ through $\alpha$. Thus finally
\begin{align}
    B_0 &= \frac{J_1}{J_2} \lambda^{-1} \nonumber \\
    &= \frac{J_1}{h_2}\frac{\Delta_{h^2}+\Delta_{J^2}\pm \sqrt{\left((h_1-h_2)^2+\Delta_{J^2}\right)\left((h_1+h_2)^2+\Delta_{J^2}\right)}}{2\Delta_{J^2}}
\end{align}
where we have defined $\Delta_{h^2} = h_1^2-h_2^2$ and $\Delta_{J^2} = J_1^2-J_2^2$. If $\Delta_{h^2}+\Delta_{J^2}>0$ we take the negative branch of the solution, and vice versa, in order to minimize $\lambda^{-1}$ and ensure convergence if possible. 

On substituting \eqref{eq:cancellation} into \eqref{eq:genfunc}, we obtain the standard geometric series generating function, immediately implying
\begin{equation}B_m = \left(\frac{J_2}{J_1}\right)^m B_0^{m+1}\label{eq:Bthreefermion}\end{equation}

As an interesting aside, although we have not made any assumptions in the derivation above as to the phase of the underlying transverse-field Ising chains, we remark here that if we do assume both chains are in the ferromagnetic phase and Taylor expand $B_0$ around $h_1=0$ and $h_2=0$, we find that the $h_1^{2n}h_2^{2k+1}$ term has numerical coefficient equal to $N(n,k)$, the number on the $n$-th row and $k$-th column of the Narayana triangle, for $n, k \geq 1$. In fact, $B_0$ is equivalent to the generating function for the Narayana numbers under the substitution $t=h_2^2/h_1^2$ and $z= -h_1^2/\Delta_{J^2}.$~\cite{Spetersen:2015}. One way of understanding this combinatorially is the following: in the perturbative construction of the SZM, after acting once with $h_2$ to obtain the lowest order possible operator string with $b_0$ present, the number of ways to act $2n$ times with the $h_1$ part of the Hamiltonian and $2k$ times with the $h_2$ part of the Hamiltonian and return to an operator string with $b_{0}$ present is exactly equal to number Dyck words of $n$ pairs of matching brackets $[$ and $]$ with $k$ nestings.

Let us now return to solving for the SZM by calculating the $A_{n,m}.$ Substituting $B_m$ from equation~\eqref{eq:Bthreefermion} into equations ~\eqref{eq:am1m} and ~\eqref{eq:a0m} yields solutions for $A_{-1,m}$ and $A_{0,m}$ as the results of partial geometric sums. To calculate $A_{n, m}$ for $n>0$, we will need to appeal to further conditions on the vanishing commutator $C=0.$ In particular, consider strings in $C$ of the form $a_{n} b_{n-1} b_{m}$ for $n>0$:
\begin{align}
    &J_2 A_{n,m+1} B_{n-1}-h_2 A_{n,m} B_{n-1}-h_2 A_{n-1,n} B_m=0 \nonumber \qquad m> n  \\
    &J_2 A_{n,n+1} B_{n-1}-h_2 A_{n-1,n} B_{n}=0 \qquad n >0 \label{eq:anm11} \\
   &\implies A_{n, m+1} = \frac{A_{n-1,n}}{B_{n-1}}\sum^m_{j=n} B_j \left(\frac{h_2}{J_2}\right)^{m-j+1}  \qquad n > 0,\, m > n.
   \label{eq:anm}
\end{align}
Substituting $B_{n}/B_{n-1} = (J_2/J_1) B_0$ and $A_{0,1}=(h_1/J_2)B_0$ from equation~\eqref{eq:a01} into equation~\eqref{eq:anm11} trivially leads to $A_{n-1,n} = \frac{h_1}{J_2}\left(\frac{h_2}{J_1}\right)^{n-1 }B_0^n$ for $n > 0$. This allows us to evaluate the sum \eqref{eq:anm} as another geometric series. Finally, we obtain the following solution for all $A_{n,m}$:
\begin{equation}
    A_{n,m} = 
    \begin{cases}
   \gamma\left[ (h_2 J_1 + B_0 \Delta_{J^2})\left(\frac{h_2}{J_2}\right)^{m}-B_0 J_1^2\left(\frac{B_0 J_2}{J_1}\right)^{m}\right], & n = -1 \\[10pt]
   \gamma \frac{h_1 B_0}{J_2} \left(\frac{h_2 B_0}{J_1}\right)^n \left[h_2 J_1 \left(\frac{h_2}{J_2}\right)^{m-n-1}-B_0 J_2^2\left(\frac{B_0 J_2}{J_1}\right)^{m-n-1}\right] , & n > -1
    \end{cases}
\end{equation}
where we have defined $\gamma= (h_2 J_1 - B_0 J_2^2)^{-1}$.

Now that we have a complete solution for $B_n$ and $A_{n,m}$, several remarks are in order. Firstly, while we have used the fact that specific terms in the commutator with the Hamiltonian $C$ vanish to fully specify the $A_{n,m}$ and $B_m$, one should, and can, check that for solutions of the above form, all terms in $C$ vanish. 
Secondly, for the SZM we require localization, such that the $A_{n,m}$ and $B_m$ are exponentially decreasing functions of $n$ and $m$. This is manifestly satisfied if the magnitudes of the ratios $\frac{h_2}{J_2}$ and $\frac{J_2 B_0} {J_1} $ are less than one. Of course, we also require that $B_0$ be real so that the SZM is Hermitian.

\begin{figure} [htbp]
\centering
 \includegraphics[width=0.6\linewidth]{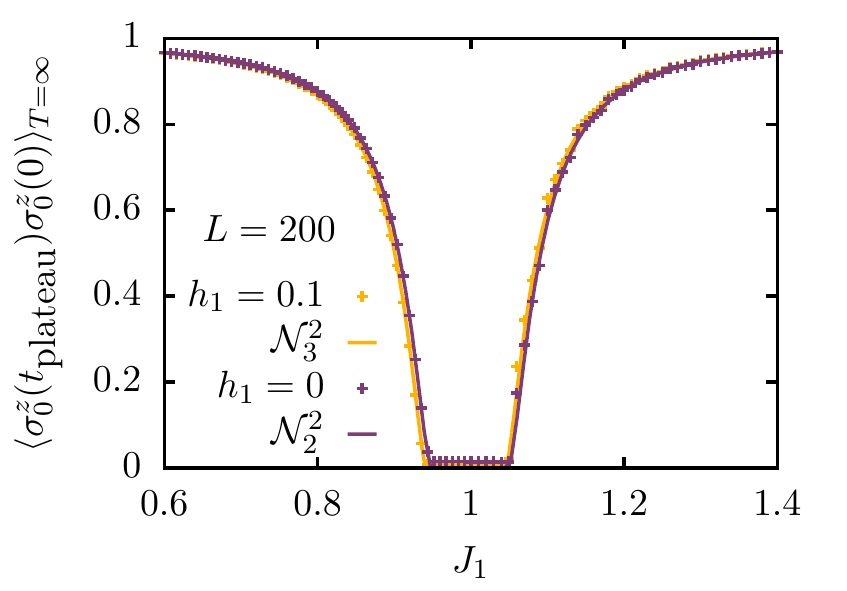}
\caption{The amplitude of the long-lived component of the autocorrelator for a toy model with a single spin at the edge. The numerical and corresponding analytical results agree precisely with equation \eqref{eq:threenorm}.}
 \label{fig:twofermion}
\end{figure}

The final calculation and condition on the exact strong zero mode is that the normalization $\mathcal{N}$ must converge. For an ansatz of the form \eqref{eq:ansatz3}, the normalization is given by
\begin{align}
\mathcal{N}_3^{-2} = \left(\sum^\infty_{l=-1} B_l^2\right)\left( \sum^
\infty_{j=-1}\sum^\infty_{k=j+1}A^2_{j,k}\right)
\end{align}
Using the explicit formss of $B_m$ and $A_{j,k}$, this sum can be easily computed as the sum of several geometric series, yielding:
\begin{equation}
\mathcal{N}_3^{-2}  =\frac{B_0 J_1^4 \left(B_0 \left(h_1^2-2
   h_2^2+J_1^2\right)+2 h_2 J_1-B_0^3 h_2^2\right)+J_1^2 J_2^2 (J_1-B_0 h_2)
   \left(J_1-\left(2 B_0^3+B_0\right) h_2\right)-B_0^2 J_2^4 (J_1-B_0
   h_2)^2}{(h_2^2-J_2^2) (J_1-B_0 h_2)^2 \left(B_0^2
   J_2^2-J_1^2\right)}
   \label{eq:threenorm}
\end{equation}
The normalization converges under the same conditions for the $A_{j,k}$ and $B_m$ remain localized. In particular, solving for the roots of the denominator which give the limit of convergence, we find:
\begin{align*}
    h_2 &= J_2, \\
    J_1 &= \pm \sqrt{J_2\frac{h_1^2-(h_2-J_2)^2}{h_2-J_2}}, \\
    J_1 &= \pm \sqrt{J_2\frac{-h_1^2+(h_2+J_2)^2}{h_2+J_2}}.
\end{align*}
For example, for $h_1=h_2=0.1$ and $J_2=1$, the normalization converges for $J_1 <0.9428\dots$ and $J_1 > 1.0444\dots$, in agreement with the numerics in Figure~5 of the main text. We can also test the exact value of the normalization~\eqref{eq:threenorm} versus the numerical plateau value, as shown in Fig.~\ref{fig:twofermion}. Here we also test the normalization $\mathcal{N}_2$ for the special case $h_1=0$ which reduces to the formula given in the main text equation (5).

\subsection{$N$ boundary spins}
Now we have the simple case of one boundary spin to act as a guide, let us turn to the full problem with $N$ boundary spins, aiming to take the limit $N \to \infty$. Recall that the zeroth order term in SZM is represented by a sea of Majorana fermion operators filling the lattice sites $j\leq 0$ for the $a$ operators and $j<0$ for the $b$ operators, due to the Jordan--Wigner string. Let us first restrict ourselves to terms in the full SZM where there are at most a single hole in the sea of the $a$ and $b$ respectively, so that there is at most only one additional $a$ operator and one $b$ operator in the right chain. We will continue using ansatz \eqref{eq:ansatz}, but simplify the notation for the single hole case by using the position of the hole in the left chain $\overline{i}$, and corresponding operator in the right chain $j$, as indices for the coefficients of the SZM operator.

Let the sequence representing the positions of the first $N$ of $N+1$ $a$-operators when there is a hole at site $\overline{i}$ be \[\mathcal{A}_{\overline{i}} = (-N,-(N-1), \dots, \overline{i-1},\overline{i+1}, \dots, 0).\] To locate all $N$ $b$-operators, we must also give the position of the displaced operator on the right chain $m$:  \[\mathcal{B}_{\overline{i} m} = (-N,-(N-1) \dots \overline{i-1},\overline{i+1}, \dots, -1, m).\]
Using these definitions, the ansatz for those terms in the SZM with at most a single hole for each fermion type becomes:
\begin{equation}
    \Psi^{\textrm{single}}_{N}=i^N\mathcal{N}\sum_{\overline{j}=-N}^{0}\sum_{\overline{n}=-N}^{-1}\sum_{\substack{k=1-\delta_{0\overline{j}} \\ m=0-\delta_{-1\overline{n}}}}^\infty \hspace{-8pt} A_{\overline{j}, k} B_{\overline{n}, m} \left(\prod_{\alpha,\beta \in \mathcal{A}_{\overline{j} } \times \mathcal{B}_{\overline{n} m} }\hspace{-15pt} a_\alpha b_\beta \right) a_{k} ,
    \label{eq:ansatzsh}
\end{equation}
For consistency with the full ansatz, we immediately set the zeroth order coefficients $A_{\overline{0},0} = B_{\overline{-1},-1} = 1$. These govern the trivial no-hole cases.

We follow the same procedure as in the single boundary-spin case. We shall demand the commutator $C = [H_\text{BI}, \Psi_N]$ vanishes, and derive recursion relations on the SZM coefficients. Of course, in general the conditions imposed on the full SZM by $C$ vanishing will relate terms in the single hole sector to those without. However, we will show that, even if we only consider terms in $C$ which solely depend on the single-hole sector, one can fully determine all $A_{\overline{i}j}, B_{\overline{i}j}$, as long one additionally demands localization to the boundary.

Recall the first step in the single boundary-spin proof was to enforce the vanishing of operator strings of the form $a_{-1} b_{-1} b_m$ and $a_{0} b_{-1} b_m$ in $C$, so that we could write the $A$ coefficients in term of the $B$. To generalise to the $N$-spin case, notice that these strings are placing a hole in the $a$-sea at $0$ and $1$ respectively and putting the left-over fermion operator at $b_m$. So for the $N$-spin case, we need to consider placing a hole at each of all $N+1$ sites in the $a$-sea; that is, we must consider operator strings of the form $\sigma^z_0 a_{\overline{n}} b_m$ for $\overline{n}\leq0$ and $m \geq 0$. 

This yields the following set of equations for $A$:
\begin{align}
A_{\overline{0},m+1}&=\frac{h_2}{J_2} A_{\overline{0},m}+\frac{J_1} {J_2}B_{\overline{-1}, m} & m\geq 0 \label{eq:an1} \\
A_{\overline{-N},m+1}&=\frac{h_2}{J_2} A_{\overline{-N},m}+\frac{h_1} {J_2}B_{\overline{-N}, m} & m\geq 0 \label{eq:an2}\\ A_{\overline{n},m+1}&=
\frac{h_2}{J_2} A_{\overline{n},m}+\frac{J_1} {J_2}B_{\overline{n-1}, m}+\frac{h_1} {J_2}B_{\overline{n}, m}  & m\geq 0 ,\, -N<n<0
\label{eq:an3}, \end{align}
where we define $A_{\overline{n},0} = 0$ for $n<0$ to cover the boundary conditions at $m=0$ of equations  \eqref{eq:an2} and \eqref{eq:an3}.   

Equations \eqref{eq:an1} and \eqref{eq:an2} are familiar from the single spin case \eqref{eq:arec3} and \eqref{eq:a01} and follow from the type of process depicted in Fig.~\ref{fig:3Ahopping}. Equation \eqref{eq:an3} is simply a combination of the two prior -- when the hole is in the `bulk' of the Jordan-Wigner string, hops from both $B$ terms are allowed. 

Recursive substitution can be used to eliminate the $A$ terms from the right hand side, yielding:
\begin{align}
    A_{\overline{0},m+1} &= \sum^m_{j=0} \frac{h_2^{m-j}}{J_2^{m-j+1}}J_1 B_{\overline{-1}, j} +\left(\frac{h_2}{J_2}\right)^{m+1}  & m \geq 0
    \nonumber \\
    A_{\overline{-N},m+1} &= \sum^m_{j=0} \frac{h_2^{m-j}}{J_2^{m-j+1}}h_1 B_{\overline{-N}, j} & m \geq 0
    \nonumber \\
     A_{\overline{n},m+1} &= \sum^m_{j=0} \frac{h_2^{m-j}}{J_2^{m-j+1}}\left(J_1 B_{\overline{n-1}, j}+h_1 B_{\overline{n}, j}\right) & m\geq 0 ,\, -N<n<0 \label{eq:aNsum}
\end{align}
These are the $N$ spin equivalents of equations ~\eqref{eq:am1m} and \eqref{eq:a0m}.

Continuing to follow the single boundary-spin proof, let us consider strings of the form $\sigma^z_0 b_{\overline{n}} a_m$ for $\overline{n}<0, m>0$, the $N$-spin equivalent of $a_{-1} a_0 a_m$ (Fig~\ref{fig:bhopping3}). These impose the conditions:
\begin{equation}
J_2 B_{\overline{n},m-1}-h_2 B_{\overline{n},m}-h_1 A_{\overline{n},m}-J_1 A_{\overline{n+1}, m}+h_2 B_{\overline{n},0} A_{\overline{n+1},m}=0 \qquad m\geq 1 \label{eq:Nbaterm}
\end{equation}
This is exactly the $N$-spin version of equation~\eqref{eq:baterm}, simply repeated $N$ times for each possible choice of hole location $\overline{n}$. Substituting in equations \eqref{eq:aNsum} we can use the same procedure which led us from equation \eqref{eq:baterm} to the full recursion relations \eqref{eq:threefermionrec} here to derive:
\begin{align}
    B_{\overline{-1},m} &=  (\alpha + \gamma B_{\overline{-1},0})B_{\overline{-1},m-1} + \beta B_{\overline{-2},m-1}-B_{\overline{-1},m-2}  & m \geq 2
    \nonumber \\
    B_{\overline{-N},m} &=  \alpha B_{\overline{-N},m-1} + \beta B_{\overline{-N+1},m-1}+\gamma B_{\overline{-N},0} B_{\overline{-1},m-1}-B_{\overline{-N},m-2}  & m \geq 2
    \nonumber \\
      B_{\overline{n},m} &=  \alpha B_{\overline{n},m-1} + \beta (B_{\overline{n+1},m-1}+B_{\overline{n-1},m-1})+\gamma B_{\overline{n},0} B_{\overline{-1},m-1}-B_{\overline{n},m-2}  & m \geq 2,\, -N<n<0, \label{eq:Nfermionrec}
\end{align}
where  $\alpha = \frac{1}{h_2 J_2} (J_2^2-J_1^2+h_2^2-h_1^2)$, $\beta = -(h_1 J_1)/(h_2 J_2)$ and $\gamma = J_1/J_2$.  This is set of $N$-coupled recursion relations for the $B_{\overline{n},m}$.

The initial conditions can also be easily derived from equations \eqref{eq:aNsum} and \eqref{eq:Nbaterm}, but before we state them, let us switch to a more natural matrix-vector notation. Let us create an $N$-dimensional column vector out of the $B_{\overline{n},m}$ at fixed $m$:
\begin{equation}
\bm{b}_m = \begin{pmatrix} B_{\overline{-N},m} \\ B_{\overline{-N+1},m} \\ \vdots \\ B_{\overline{-1},m}
\end{pmatrix}
\end{equation}

Now we can consider matrices acting on this vector: let us define a tridiagonal Toeplitz matrix $T$, a square matrix $\Gamma$ with all but the last column zero, and their sum $F$:
\begin{equation}
T = \begin{pmatrix}
\alpha & \beta &  \\
\beta  & \alpha & \beta  \\
& \beta & \ddots & \ddots \\
& & \ddots & \ddots & \beta \\
& & & \beta & \alpha
\end{pmatrix}_{N\times N}
\qquad \Gamma = \begin{pmatrix}
0 & 0 &\dots & \uparrow\\
\vdots  & \vdots &\dots &\gamma \bm{b}_0  \\
0& 0 & \dots & \downarrow 
\end{pmatrix}_{N\times N} \qquad   F = T + \Gamma
\label{eq:Fdef}
\end{equation}
Then in a $2N$-dimensional vector space, let us further introduce the block matrix $\mathcal{M}$ and block vector $\mathfrak{b}_n$:
\begin{equation}
\mathcal{M}= \begin{pmatrix}
0 & I   \\
-I  & F 
\end{pmatrix}_{2N \times 2N} \qquad \mathfrak{b}_n = \begin{pmatrix}
    \bm{b}_n \\
    \bm{b}_{n+1}
\end{pmatrix}, 
\end{equation}
where $I$ is the $N\times N$ identity matrix.

Finally the coupled recurrence relations \eqref{eq:Nfermionrec} can be written as a single matrix equation
\begin{equation}
    \mathfrak{b}_m = \mathcal{M}^m \mathfrak{b}_0
    \label{eq:blockrecurrence}
\end{equation}
The initial conditions are given by
\begin{equation}
\mathfrak{b}_0 = \begin{pmatrix}
    \bm{b}_0 \\
    \bm{b}_{1}
\end{pmatrix} =   \begin{pmatrix}
    \bm{b}_0 \\
    \bm F{b}_{0} - \gamma \bm{e}_n
\end{pmatrix},
\end{equation}
where $\bm{e}_n^{\intercal}=(0 \dots 1)$ is the $N$-th unit vector.

Notice the terms in $\bm{b}_m$ are coefficients of strings in the SZM which move further and further from the boundary as $m$ increases. Thus to have localization, we must at least require that all the eigenvalues of the eigenvectors of $\mathcal{M}$ which $\mathfrak{b_0}$ has overlap with be less than one. 

In fact the block diagonal structure of $\mathcal{M}$ makes it easy to calculate its the eigenvalue $\lambda_i$ in terms of the eigenvalues $f_i$ of the matrix $F$:\begin{equation}f_i = \frac{\lambda_i^2+1}{\lambda_i}\label{eq:palind}.\end{equation} By replacing $\lambda_i$ with $\lambda_i^{-1}$ in the above equation, it is clear that if $\lambda_i$ is an eigenvalue $\mathcal{M}$ so is $\lambda_i^{-1}$. Thus if all the eigenvalues of $\mathcal{M}$  are real and not equal to one, they split into pairs of magnitude less than and greater than one respectively. This should be reminiscent from the single boundary spin case, where we found the characteristic polynomial \eqref{eq:genfunc} had two reciprocal roots, and we chose $B_0$ to cancel the root with magnitude greater than one. Here we must instead chose $\mathfrak{b_0}$ such that it lies in the $N$-dimensional subspace spanned by the eigenvectors of $\mathcal{M}$ with eigenvalues magnitude less than one. This will provide us with $N$ further constraints, fixing the remaining $N$ unknown SZM coefficients $\bm{b}_0$.

Before we can do this, we must first show that the eigenvalues of $\mathcal{M}$ are indeed real and not equal to one. Solving  equation \eqref{eq:palind} for $\lambda_i$ we find:
\begin{equation}
\lambda_i = \frac{1}{2} \left(f_i \pm \sqrt{f_i^2-4}\right)
\label{eq:Meigs}
\end{equation}
Hence the required conditions are satisfied if all the $f_i$ are real and $|f_i| > 2$.

Recall from equation~\eqref{eq:Fdef} that the matrix $F$ is the sum of a tridiagonal Toeplitz matrix $T$ and a matrix with only one non-zero column $\Gamma$, which depends on the initial conditions $\bm{b}_0$. Let us first consider only $T$, for which the eigensystem is well known. We will sketch the derivation in a manner which will be illuminating later. Let $\bm{v}^{\intercal} = (v_1 v_2 \dots v_N)$ be an eigenvector such that $T \bm{v} = t \bm{v}.$ Then the elements of $\bm{v}$ satisfy the recursion relation:
\begin{equation}
    \beta v_{n-1} + \alpha v_n + \beta v_{n+1} = t v_n
    \label{eq:trec}
\end{equation}
with boundary conditions $v_0=v_{N+1}=0$. The characteristic polynomial of the recursion has roots:
\[r_{\pm} = \frac{1}{2\beta} \left( \alpha-t \pm \sqrt{(\alpha-t)^2-4\beta^2}\right)\]
Let us reparamaterise $t$ with $\theta$ via
    $t = \alpha + 2 |\beta| \cos{\theta}$.
Then $r_{\pm} = e^{\pm i \theta}$. The general solution to the recursion relation is thus $v_n = c_+ e^{i n\theta} + c_-  e^{-i n\theta}$. The boundary condition $v_0=0$ gives $c_+=-c_-$, which combined with the other boundary condition $v_{N+1}= 0$ implies $\sin{(N+1)\theta} = 0$. Finally then the eigenvalues are
\begin{equation}
    t_n = \alpha + 2 |\beta| \cos{\left(\frac{\pi n}{N+1}\right)} \qquad n \in [1, N].
\end{equation}
If we neglect the effect of $\Gamma$ and let $f_n=t_n$, then the condition for the localization of the SZM becomes:
\begin{align}
|\alpha \pm 2 \beta | &> 2 \label{eq:condition} \\
\left|\frac{(J_2^2-J_1^2+h_2^2-h_1^2)\pm 2 h_1 J_1}{h_2 J_2} \right| &>2
\end{align}
Let $ \sgn(\alpha \pm 2 \beta)= \pm_{1}1$. Then the limits on localization is given by
\begin{align}
(J_2^2-J_1^2+h_2^2-h_1^2)\pm 2 h_1 J_1 \mp_{1} 2 h_2 J_2&=0 \\
(J_2 \mp_{1} h_2)^2 &= (J_1 \mp h_1)^2.
\end{align}
Notice that $(J_i\pm h_i)$ yield the maximum and minimum energies of an infinite TFIM chain with Ising coupling $J_i$ and transverse field $h_i$. Thus the condition on localization of the SZM represents exactly the condition that the bands of the disconnected chains do not overlap, as seen in the numerics in Fig.~4 of the main text and as expected from the simple argument presented there.

Of course, before we may draw any conclusions, we must examine the effect of the perturbation of $F$ by $\Gamma$. Suppose we have eigenvector of $F$ such that $F\bm{v} =f \bm{v}$. Then just like for $T$ we can write down a recursion relation for the elements of $\bm{v}$. For notational simplicity let us write the $n$-th element of $\bm{b}_0$ as $\tilde{b}_n$ (not to be confused with the Majorana fermion operator $b_n$). Then the recursion relation becomes:
\begin{equation}
    \beta v_{n-1} + \alpha v_n + \beta v_{n+1}+\gamma v_N \tilde{b}_n = f v_n.
\end{equation}

Let us rescale $\bm{v}$, defining a new vector $\bm{y}=\bm{v}/v_N$. This amounts to a choice of normalisation as long as $v_N \neq 0$. Then the recursion relation becomes:
\begin{equation}
    \beta y_{n-1} + \alpha y_n + \beta y_{n+1}+\gamma \tilde{b}_n = f y_n,
    \label{eq:frec}
\end{equation}
with boundary conditions $y_0=0, y_{N}=1, y_{N+1}=0$. Notice this is the same recurrence relation as for $T$ (eqn. \ref{eq:trec}), but with different boundary conditions, and inhomogeneous as opposed to homogeneous. The general solution is thus the same; let $f = \alpha + 2 |\beta| \cos{\theta}$, then $y_n^\mathrm{gen} = c_+ e^{i n\theta} + c_-  e^{-i n\theta}$. We must add the particular solution to deal with the $\gamma b_n$ term; from the method of variation of parameters we find:
\begin{equation}
    y_n^\mathrm{part} =  -\frac{\gamma}{\beta} \sum_{j=1}^{n-1}\frac{e^{i(n-j)\theta}-e^{-i(n-j)\theta}}{e^{i\theta}-e^{-i\theta}} \tilde{b}_j.
    \label{eq:ypart}
\end{equation}
We can now apply the boundary conditions the full solution $y_n = y_n^\mathrm{gen} + y_n^\mathrm{part}$. From $y_0=0$ we still find $c_+=-c_-$. Then $y_N=1$ implies
\begin{equation}
    c_+ =\frac{1}{e^{iN\theta}-e^{-iN\theta}}\left(1+\frac{\gamma}{\beta} \sum_{j=1}^{N-1}\frac{e^{i(N-j)\theta}-e^{-i(N-j)\theta}}{e^{i\theta}-e^{-i\theta}} \tilde{b}_j\right).
\end{equation}
Finally substituting these results into $y_{N+1}=0$ yields
\begin{align}
    \frac{e^{i(N+1)\theta}-e^{-i(N+1)\theta}}{e^{iN\theta}-e^{-iN\theta}}\left(1+\frac{\gamma}{\beta} \sum_{j=1}^{N-1}\frac{e^{i(N-j)\theta}-e^{-i(N-j)\theta}}{e^{i\theta}-e^{-i\theta}} \tilde{b}_j\right)&=\frac{\gamma}{\beta} \sum_{j=1}^{N}\frac{e^{i(N-j+1)\theta}-e^{-i(N-j+1)\theta}}{e^{i\theta}-e^{-i\theta}} \tilde{b}_j\nonumber \\
    \frac{e^{i(N+1)\theta}-e^{-i(N+1)\theta}}{e^{iN\theta}-e^{-iN\theta}}&= \frac{\gamma}{\beta}\left(\sum_{j=0}^{N-1}\frac{e^{i(N-j)\theta}-e^{-i(N-j)\theta}}{e^{i\theta}-e^{-i\theta}} [\tilde{b}_{j+1} - \tilde{b}_j ]\right),
    \label{eq:fullNev}
\end{align}
where to simplify notation we set the fictitious $\tilde{b}_0=0$. If we knew $\bm{b}_0$ we could now solve equation~\eqref{eq:fullNev} for the $N$ valid solutions for $\theta$ in the same way we solved $\sin(N+1)\theta$ to find the eigenvalues of $T$.

Recall that we may fix $\bm{b}_0$ from the condition that $\mathfrak{b_0}$ must lie in the $N$-dimensional subspace spanned by the eigenvectors of $\mathcal{M}$ with eigenvalues magnitude less than one. Explicitly, let $\lambda_i$ with $|\lambda_i| > 1$ be the eigenvalue of $\mathcal{M}$ related to $f_i$ via equation \eqref{eq:Meigs}. Recall that $\lambda_i^{-1}$ is then also an eigenvalue. The corresponding eigenvectors of $\mathcal{M}$ are then related to the eigenvectors $\bm{y}_i$ of $F$ by:
\begin{equation}
    \mathfrak{y}_{i}^{\pm} = \begin{pmatrix}
    \bm{y}_i \\
    \lambda_i^{\pm 1}\bm{y}_{i}
\end{pmatrix}
\end{equation}
The localization condition is $\mathfrak{b_0}\cdot \mathfrak{y}_i^{+} = 0$ for all $i$, or
\begin{equation}
\bm{y}_i \cdot \left[(1+ \lambda_i F) \bm{b}_0 - \gamma \lambda_i \bm{e}_n\right]  =0 \qquad \forall i
\label{eq:orthog}
\end{equation}
These conditions combined with the full solution for $\bm{y}_i$ and $f_i$ in equations \eqref{eq:ypart} and \eqref{eq:fullNev} in terms of $\bm{b}_0$ in theory allow us to fully determine $\bm{b}_0$ as we did for $B_0$ in the single boundary spin case. Unfortunately, they are a set of complicated coupled non-linear equations.

Instead of directly solving them, let us remark that in order to show localization we need only bound $f_i > 2$. In fact the specific form of $f_i = \alpha + 2 |\beta| \cos{\theta_i}$ makes this particularly easy: as long as the $\theta_i$ are not complex, but either real or purely imaginary, the condition is certainly satisfied if $|\alpha\pm 2 \beta| > 2$. This is the same localization condition \eqref{eq:condition} we had before we included $\Gamma$ into $F$! 

Let us examine equation \eqref{eq:fullNev} and show that the solutions $\theta_i$ are indeed either real or purely imaginary. Firstly, notice that the $\tilde{b}_j = B_{\overline{j-N-1},0}$ must themselves decay as $j$ decreases in order to ensure localization. In fact it is possible to show this follows under the same condition \eqref{eq:condition} that all the other terms in the SZM are localized, by going through the all the steps outlined above in this section but for the problem where there are $N$-boundary spins on the right chain, rather than the left. In the fermionic language, this is the problem of a finite number of holes hopping into an infinite filled sea, rather than reverse we have been considering. In the limit $N\to \infty$, they of course must yield the same answer. For that version of the problem, one may derive an equivalent recurrence relation to \eqref{eq:blockrecurrence} with a different $\mathcal{M^\prime}$, but whose eigenvalues $\lambda^\prime$ separate into pairs magnitude less than and greater than one under the same conditions as $\mathcal{M}$.   $\bm{b}_0$ are not the initial conditions, but instead each element follows from the next step in the recurrence relation. Thus assuming the conditions are met, the $b_j$ must decay at least as fast as powers of the the largest eigenvalue of $\mathcal{M^\prime}$ with magnitude less than 1; or equivalently $|b_j| < |b_N e^{(j-N)\xi}|$ for some $\xi>0$.

To gain intuition on equation \eqref{eq:fullNev} let us consider the simple limit $\xi \to \infty$; i.e. $b_j = 0$ for $j\neq N$. Then if $\theta$ is real equation \eqref{eq:fullNev} simpilifies to:
\begin{equation}
    \sin{(N+1) \theta} = d \sin{N \theta},
    \label{eq:sinN}
\end{equation}
for $d = \gamma b_N/\beta$. We require unique, non-trivial solutions for $f_i$ and $\bm{y}_i$, so we restrict the domain to $0 <\theta < \pi$. In this domain, it is clear that there are $N$ solutions to equation \eqref{eq:sinN} if $|d|<(1+1/N)$, because the sine wave on the right hand side is both lower frequency and grows more slowly than the one on the left at $\theta=0$. The solutions are perturbations of the solutions to $\sin{(N+1) \theta}=0$, i.e. the $\theta_i$ associated with the eigenvalues $t_i$ of $T$. In fact in the limit $N\to \infty$, these solutions become dense over the domain and it is manifest (and easy to show in perturbation theory) that $f_i \to t_i \,\, \forall i$ as long as $|d|<1$.

On the other hand, if $|d|>(1+1/N)$ we `lose' the largest of these solutions, because the lower frequency, higher amplitude sine wave completely skips over the first zero of the other. Instead, we must consider pure imaginary $\theta$:
\begin{equation}
\sinh{i(N+1)\theta} = d\sinh{iN\theta}.
\label{eq:sinhN}
\end{equation}
Indeed by comparing the gradient of the left versus right hand sides of the above equation at $\theta=0$, it is manifest that there are two non-trivial pure imaginary solutions if and only if $|d|>(1+1/N)$. Because these two solutions necessarily have the form $\theta_\pm = \pm \theta_c$, they correspond to a single unique eigenvalue $f_i=\alpha + 2 |\beta| \cos{\theta_i}$ of $F$.

Thus we have shown that all $N$ eigenvalues are accounted for by pure real or imaginary $\theta$, and thus the condition for localization\eqref{eq:condition} holds, if $\xi \to \infty$. Let us now allow the $\bm{b}_0$ to have an exponential tail by requiring merely $\xi> 0$. Then for the slowest possible decay of the $\bm{b}_0$, equation \eqref{eq:sinN} becomes:
\begin{equation}
\sin{(N+1)\theta} = d\sin{N\theta}\left(\sum_{j=0}^{N-1}\frac{\sin{(N-j) \theta}}{\sin\theta} [e^{(j+1-N)\xi}- e^{(j-N)\xi}] +\frac{\sin{N \theta}}{\sin\theta} e^{-N\xi}\right),
\end{equation}
where the last term on the right hand side, which is irrelevant in the limit $N\to \infty$, technically must appear because we defined $\tilde{b}_0=0$.

The bracketed term is a simple geometric series in $j$ (which may be made more obvious by leaving the sine in the numerator as the sum of exponentials). This may be calculated using the usual formula in the limit $N\to \infty$:
\begin{equation}
\sin{(N+1)\theta} = d\sin{N\theta}\left(\frac{e^\xi-1}{2\cosh \xi - 2\cos \theta}\right),
\label{eq:bracketedreal}
\end{equation}
For small positive $\xi$, the function in brackets is strongly peaked about $\theta=0$ before monotonically decaying in the domain $0 < \theta < \pi$, never falling below zero. As $\xi$ increases the peak at $\theta=0$ falls and the function approaches the constant 1, recovering the limit considered above. It is thus clear that the bracketed term in equation \eqref{eq:bracketedreal} does not qualitatively change the behaviour of the solutions compared to equation \eqref{eq:sinN}: there are still either $N$ or $N-1$ real roots, although now the condition for an imaginary root depends on both $d$ and $\xi$.

Similarly the pure imaginary roots,
\begin{equation}
\sinh{i( N+1)\theta} = d\sinh{i N\theta}\left(\frac{e^\xi-1}{2\cosh \xi - 2\cosh i \theta}\right),
\label{eq:bracketedimag}
\end{equation}
have the same structure as for equation \eqref{eq:sinhN}: there are two pure imaginary solutions $\theta = \pm \theta_c$, leading to one unique solution for $f_i$, under the same conditions that there are only $N-1$ real roots. (Notice it might appear that there are always a further two solutions associated with the divergence of the bracketed term as $i \theta \to \xi$; however, this is an artefact of taking the limit $N\to \infty$ in the brackected term and not for the hyperbolic sines -- the geometric sum does not diverge except in this limit, and it is clear that the left hand side still diverges faster than the right for finite fixed $\theta$ under this limit.)

We have shown that the solutions to $\theta$ are always either pure real or imaginary. This means that the eigenvalues of $F$, $f_i = \alpha + 2 |\beta| \cos{\theta_i}$ still satisfy  $f_i > 2$ under condition \eqref{eq:condition}. Thus if this condition holds, the $\bm{b}_m$, i.e. $B_{\overline{n},m}$, are exponentially localized to the boundary. In order to prove the full localization of the SZM, we must consider the $A_{\overline{n},m}$, as well as all the $A$ and $B$ terms outside of the single hole sector. This follow exactly the same procedure as in the single boundary spin case where we calculated the $A_{n,m}$ (in both single and two hole sectors) from the $B_m$ (eqn~\eqref{eq:anm}).

Firstly,  for the single-hole sector, it is immediately apparent from equation~\eqref{eq:aNsum} that the $A_{\overline{n},m}$ will be localized if the $B_{\overline{n},m}$ are localized, and additionally $|h_2| < |J_2|$. Secondly, to treat the multiple-hole sectors, we must consider the vanishing of additional operator strings in the commutator $C$. For example, we can consider terms in $C$ of the form  $\sigma^z_0 b_{\overline{-2}} a_{\overline{0}} b_{0}b_j$ for $j > 0$ to calculate the two-hole coefficients $B_{\overline{-2},\overline{-1}, 0, j}$, in terms of the single-hole sector coefficients. Once we have the $B_{\overline{-2},\overline{-1}, 0, j}$ we can work forwards by replacing the operator at 0 with 1 and considering the resultant string in $C$, and so on to arbitrary $n$; simultaneously, we can work backwards by moving the holes at $-1$ and $-2$ to the left. Once the two-hole $B$ terms have been determined, we can determine the $A$ by considering strings of the form $\sigma^z_0 a_{\overline{n}} a_{\overline{n+1}} b_{\overline{n}} a_j b_{j-1} b_l$ in $C$ for $\overline{n}<0, \overline{m}<n, j>0, l>=n$.

This allows us to calculate all the two-hole sector coefficients in terms of the single hole sector coefficients. Because all the terms which contribute to a single operator string in $C$ must be related by at most two hops of fermion operators (conserving $a$ and $b$ number), it is always possible to write an $n$-hole sector in terms of coefficients of the $n-1$ and $n-2$ hole sectors, and thus by induction the single-hole sector. Once we have the result in terms of the single-hole coefficients, we can expand them in terms of powers of the eigenvalues of $\lambda_i$ of $\mathcal{M}$ using equations~\eqref{eq:aNsum} and \eqref{eq:blockrecurrence}. For all cases in the two-hole sector, the result is much as for the two-hole case for the single boundary-spin \eqref{eq:anm}: the coefficients are the sum of products of different pairs of powers of $\lambda_i^m$ and $(h_2/J_2)$, divided by a sum of powers of $\lambda_i$ and $h_2/J_2$. The powers in the numerator and denominator are such that we again we find exponential localization, if we assume additionally that the $\bm{b}_0$ are localized, which as we have argued above occurs under the same conditions as $\lambda_i <1$.

This is not a coincidence: if all $(m<= n)$-hole sectors are exponentially-localized (for $n>0$), the $(n+1)$-sector must be if all the coefficients in front of the operator strings in $C$ vanish. We can see this by the following simple argument: take the coefficient of an arbitrary operator string $S$ in the $(n+1)$-hole sector. We may relate it to the coefficients of the $(m<n)$-hole sectors by considering vanishing of the appropriate term in $C$. Now translate the leftmost cluster of adjacent fermion operators in $S$ by $j$ sites to the left. Due to the translation symmetry of the Hamiltonian in the left chain, the coefficient of the new string $S^\prime$ is related to the $m$-hole sector by the \textbf{same} equation, but with the $n$-hole sector coefficients replaced by those $j$ powers of the $\lambda_i$ or $h_2/J$ smaller than for $S$, by the assumption of exponential localization of the $m$-hole sectors. Thus if the coefficient of $S^\prime$ is not exponentially localized compared to $S$, the two equations cannot be consistent. We can repeat this procedure with the second leftmost cluster of fermion operators and so forth, and may also make the same argument concerning the translation of clusters of holes to the left (this time utilising the translational symmetry of the chain on the right).

Given that rearrangement of the $(n+1)$-hole sector must be consistent with exponential localization, we now need only show that there exists a process which takes the $n$-hole sector to the $(n+1)$-hole sector for which the coefficient in the $(n+1)$-hole sector is localized. To do this, consider strings in $C$ of the form $\sigma^z_0 (\prod_{j=-n}^{-1} a_{\overline{j}} b_{\overline{j}}) (\prod_{j=0}^{n-1} a_j b_j) b_{n}$; Because the holes and corresponding fermion operators have been placed in a block on the left and right of the boundary respectively, the possible fermion operator hops out of this string are limited, and thus so are the number of coefficients in the SZM which can commute with the Hamiltonian to generate this string. We find 
\begin{align}
B_{\overline{[-n-1,-1]}; [0,n]}=-\frac{1}{J_1 A_{\overline{[-n+1,-1]};[0,n-1]}} \bigg(&B_{\overline{[-n,-1]}; [0,n-1]}\big(h_2 A_{\overline{[-n,-1]};[0,n]}-J_2 A_{\overline{[-n,-1]};[0,n-1] \bigcup \{n+1\}}\big) \\ &+ J_2 B_{\overline{[-n,-1]}; [0,n-2]\bigcup \{n\}} A_{\overline{[-n,-1]};[0,n]}\bigg),
\end{align}
where the set of integers with the overline before the semicolons in the subscripts of the SZM coefficients are the indices of the holes in the left hand side of the chain, while those after are the indices of the fermion operators in the right. The SZM coefficient on the left hand side is in the $(n+1)$-hole sector, while those on the right are in the $n$-hole sector. If we assume exponential localization of the latter, then simple power counting immediately yields that the left hand side coefficient must be suppressed by the correct power of the $\lambda_i$ and $h/J$. Putting all these arguments together, we may conclude that the $(n+1)$-hole sector is indeed localized if those sectors with fewer holes also are.

In summary, we have shown that an operator which commutes with the Hamiltonian and is exponentially localized to the boundary exists if and only if the following  conditions are satisfied:
\begin{equation}
\left|\frac{(J_2^2-J_1^2+h_2^2-h_1^2)\pm 2 h_1 J_1}{h_2 J_2} \right| >2, \qquad \left|\frac{h_2}{J_2}\right|< 1
\label{eq:allconditions}
\end{equation}
To prove that the exact SZM exists, we must also show that its normalization converges. For SZMs of the form described by our ansatz \eqref{eq:ansatz}, this means the sum of squares of the coefficients must converge. We have just determined that the coefficients of terms $n$-hops away from the zeroth-order term $\sigma^z_0$ is exponentially small in $n$; however, recall that we have also shown in equation ~\eqref{eq:countingterms} that the numbers of such terms is subexponential! Thus the normalization must indeed converge. Finally, we can conclude that an exact SZM exponentially localized to $\sigma^z_0$ exists if the conditions \eqref{eq:allconditions} are satisfied.

\subsection{Exact form of the SZM for $N$ boundary spins and the Narayana numbers}

Although we were unable to solve equations\eqref{eq:ypart}, \eqref{eq:fullNev}, and \eqref{eq:orthog} for the SZM coefficients, we may gain some insight into the form of the solution from perturbation theory. In particular, we can show that it is extremely unlikely that the coefficients can be written as elementary functions of the couplings and appear to be related to the generating functions of the Narayana numbers.

By inspection of the perturbation theory up to 15th order, it appears that for the ferromagnetic-ferromagnetic SZM, the simplest non-trivial SZM coefficient is:
\begin{equation}
B_{\overline{-1}, 0} = \frac{J_1 h_2}{\Delta_{J^2}}\left(1+ \frac{J_2^2 h_1^2}{\Delta_{J^2}}\sum_{n=1}^\infty\sum_{j, k=0}^{n-1} \frac{ J_1^{2(n-j-1)} J_2^{2j} h_1^{2(n-k-1)} h_2^{2k}}{\Delta_{J^2}^{2n}}N(n,k) N(n,j)\right),
\end{equation}
where recall $N(n, k)$ are the Narayana numbers and $\Delta_{J^2} = J_1^2-J_2^2$. Thus where $B_0$ in the single boundary spin case was simply the generating function of the Naryana numbers, in the $N \to \infty$ case the equivalent $B_{\overline{-1}, 0}$ seems to be the termwise (or Hadamard) product of \emph{two} Narayana generating functions. As far as we are aware, this has not been calculated in the literature. However, the generating function of the closely related squares of the Catalan numbers has been calculated, and it turns out to be an elliptic integral~\cite{Slando_plane_1993}.

\section{Operator perturbation theory}
\subsection{Background and proof of SZM construction}
Earlier we gave the intuition that resonances occur when there is an energy conserving process to flip the degree of freedom associated with the 0$^\text{th}$ order term of the almost SZM using terms from the perturbative part of the Hamiltonian.
However, there were other resonances we could imagine existing based on this reasoning that we don't actually observe.
In the boundary Ising model, for instance, if in addition to flipping spin $-1$ with $-J_1\Z_{-1}\Z_0$ and spin $0$ with $-h_2\X_0$ we also flipped spins $-3$ and $-2$ with $-J_1\Z_{-3}\Z_{-2}$, we could have an energy conserving process at $h_1=J_2/3$ coming in at third order, but no such term exists.

Here we lay out another way to construct almost SZMs (or SZMs) that is helpful for gaining insight into why certain poles appear or do not appear in a given mode.
This construction uses an operator perturbation theory; the intuition is that because the (almost) SZM (approximately) does not evolve in time, we should be able to write it as a linear combination of operators that are stationary under evolution by the Liouvillian (i.e. the commutator with the Hamiltonian).
When we perturb the Hamiltonian, and thus the Liouvillian and its stationary operators, the corresponding linear combination of perturbed operators should still be approximately stationary.
We can think of this in analogy to ordinary perturbation theory on a system's eigenstates.
Here instead we're concerned with eigen\emph{operators} of the Louivillian, which is a linear superoperator (i.e. a map from operator space to itself).

We will utilize the operator perturbation theory derived in Ref.~\cite{SPhysRevA.12.2549} to construct SZMs.
First we will restate the relevant definitions.
We consider perturbing a Hamiltonian $H_0\rightarrow H_0+\lambda H_1$, where we know how to diagonalize $H_0$,
\begin{equation}
H_0|n^{(0)}\rangle=E_n^{(0)}|n^{(0)}\rangle.
\end{equation}
We will also define the Liouville superoperators
\begin{align}
    L_0=[H_0,\cdot]\\
    L_1=[H_1,\cdot]\\
    L=[H,\cdot].
\end{align}

Analogously to applying perturbation theory to wavefunctions, Ref.~\cite{SPhysRevA.12.2549} describes how operators change in the Heisenberg picture under the effect of the perturbation.
To do this, we need some set of unperturbed basis operators $\{R_n^{(0)}\}$ that are eigenoperators of $L_0$:
\begin{equation}
    L_0R_n^{(0)}=\omega_n^{(0)}R_n^{(0)}\label{opeig}
\end{equation}
such that
\begin{equation}
    L_0\omega_n^{(0)}=0\label{weig},
\end{equation}
where generically the eigenvalues $\{\omega_n^{(0)}\}$ are themselves allowed to be operators.
In particular, one choice that will work for any $H_0$ is $\{R_{ij}^{(0)}\}$, with
\begin{equation}
    R_{ij}^{(0)}=|i^{(0)}\rangle\langle j^{(0)}|
\end{equation}
and 
\begin{equation}
    \omega_{ij}^{(0)}=E_i^{(0)}-E_j^{(0)},
\end{equation}
as can be readily verified using Eq.~\eqref{opeig}.

Finally, we define perturbed versions of the eigenoperators and eigenvalues, and demand that they obey analogous conditions to Eqs.~\eqref{opeig} and \eqref{weig}:
\begin{align}
    S_n=&\sum_m \lambda^n R_n^{(m)}\\
    \Omega_n=&\sum_m \lambda^n \omega_n^{(m)}
\end{align}
such that
\begin{align}
    LS_n=&\Omega_nS_n\\
    L\Omega_n=&0.
\end{align}
With these definitions out of the way, we will decompose the unperturbed SZM into the $\{R_{ij}^{(0)}\}$ basis,
\begin{equation}
    \Psi^{(0)}=\sum_{i,j}c_{ij}R_{ij}^{(0)}
\end{equation}
and prove that by replacing each $R_{ij}^{(0)}$ with $S_{ij}$ we get the SZM in the perturbed system (or an almost SZM, truncating appropriately):
\begin{equation}
    \Psi=\sum_{i,j}c_{ij}S_{ij}.
\end{equation}
as long as the perturbation does not break the symmetry associated with the SZM.

\noindent\emph{Proof.} Let $\psi=\sum_{i,j}c_{ij}S_{ij}$. Then acting with $L$ we have
\begin{equation}
    L\psi=\sum_{i,j}c_{ij}LS_{ij}=\sum_{i,j}c_{ij}\Omega_{ij}S_{ij}.
\end{equation}

Now, since
\begin{equation}
\begin{split}
    \Psi^{(0)}=&\sum_{ij}|i^{(0)}\rangle\langle i^{(0)}|\Psi^{(0)}|j^{(0)}\rangle\langle j^{(0)}|\\
    =&\sum_{ij}\langle i^{(0)}|\Psi^{(0)}|j^{(0)}\rangle R_{ij}^{(0)},
\end{split}
\end{equation}
we have that $c_{ij}=\langle i^{(0)}|\Psi^{(0)}|j^{(0)}\rangle$. 
Furthermore, we know that, up to corrections exponentially suppressed in $L$, $\Psi^{(0)}$ takes each eigenstate of $H_0$ to a state in another sector of the associated symmetry~\cite{SKemp_2017}. 2
These pairs (or, generally, groups of $m$ states for SZMs with $\Psi^m=1$) will also be degenerate up to exponentially small corrections in $L$, and if the pertubation doesn't break the SZM's symmetry that degeneracy will be preserved.

Then since for this basis $\omega_{ij}^{(n)}=E_{i}^{(n)}-E_{j}^{(n)}$\cite{SPhysRevA.12.2549}, we have $\Omega_{ij}=E_{i}-E_{j}$. So for any $i,j$, either $c_{ij}$ or $\Omega_{ij}$ is exponentially small in $L$. 
Thus $\norm{L\psi}=\norm{\sum_{i,j}c_{ij}\Omega_{ij}S_{ij}}<e^{-\alpha L}$ for some constant $\alpha$, and $\psi$ is an SZM. $\square$
\subsection{Operators SZM construction for the boundary Ising model}\label{PTconstruction}
Now we will apply operator perturbation theory to reconstruct the almost SZM for the boundary Ising model to second order.
In doing so, we will be a little more general and not specify $H_0$ other than requiring that it has an unperturbed SZM $\Psi^{(0)}=\sigma_0^z$ and the eigenstates we will specify below.
We will, however, specify $V$:
\begin{equation}
    V_\text{BI}=-J_1\sum_{j=-\infty}^{-1}\sigma_j^z\sigma_{j+1}^z
    -h_2\sum_{j=0}^\infty\sigma_j^x.
\end{equation}
We will also work in the basis of states
\begin{equation}
    \left\{\left(\bigotimes_{\alpha=-\infty}^{-1}|\leftrightarrow_\alpha\rangle\right)\otimes
    \left(\bigotimes_{\beta=0}^{\infty}|\updownarrow_\beta\rangle\right)\right\}
    \label{basis}
\end{equation}
where $|\leftrightarrow_\alpha\rangle$ is an eigenstate of $\X_\alpha$ and $|\updownarrow_\beta\rangle$ is an eigenstate of $\Z_\beta$.
In particular, these are eigenstates for $H_\text{BI}[J_1=h_2=0]$.
Finally, for simplicity we will adopt some notation. First, in the expressions that follow we will drop $(0)$ superscripts for the unperturbed states and energies.
Indices $i$ and $j$ will refer to elements of the basis Eq.~\eqref{basis}, whereas $\alpha$ and $\beta$ will refer to sites. A prime on a Greek letter or a number, e.g.~$\alpha^\prime$, will refer to flipping that spin, and a double prime refers to flipping a spin as well as the spin to the right (so $\beta^{\prime\prime}$ refers to flipping spins $\beta$ and $\beta+1$).

For the $R_{ij}$ basis we're using, corrections to a basis operator can be written in terms of the perturbative corrections to the eigen\emph{states}\cite{SPhysRevA.12.2549}:
\begin{equation}
    R_{ij}^{(n)}=\sum_{m=0}^n|i^{(m)}\rangle\langle j^{(n-m)}|.
\end{equation}
We then note that since $\Psi^{(0)}=\sigma_0^z$ is diagonal in this basis, we only need corrections to $R_{ii}$:
\begin{equation}
    \Psi^{(0)}=\sum_{ij}\langle i|\sigma_0^z|j\rangle R_{ij}^{(0)}=\sum_{i}\langle i|\sigma_0^z|i\rangle R_{ii}^{(0)},
\end{equation}
and therefore
\begin{equation}
    \Psi^{(n)}=\sum_{i}\langle i|\sigma_0^z|i\rangle R_{ii}^{(n)}.\label{ZPT}
\end{equation}
The first and second order corrections are
\begin{equation}
    R_{ii}^{(1)}=-h_2\sum_{\alpha\ge 0}\frac{\ketbra{i;\ab}{i}+\text{h.c.}}{E_i-E_i^{\ab}}
    -J_1\sum_{\alpha<-1}\frac{\ketbra{i;\abb}{i}+\text{h.c.}}{E_i-E_i^{\abb}}
    -J_1\bra{i}\Z_0\ket{i}\frac{\ketbra{i;-1^\prime}{i}+\text{h.c.}}{E_i-E_i^{-1^\prime}}
\end{equation}
and
\begin{equation}
\begin{split}
    R_{ii}^{(2)}=&
    h_2^2\bigg(\sum_{\substack{\alpha,\beta\ge 0 \\ \alpha\neq\beta}}\frac{\ketbra{i;\ab\bb}{i}+\text{h.c.}}{(E_i-E_i^{\ab})(E_i-E_i^{\ab\bb})}-\sum_{\alpha\ge 0}
    \frac{\ketbra{i}{i}}{(E_i-E_i^{\ab})^2}+\sum_{\alpha,\beta\ge 0}\frac{\ketbra{i;\ab}{i;\bb}}{(E_i-E_i^{\ab})(E_i-E_i^{\bb})}
    \bigg)\\
    &+J_1^2\bigg(\sum_{\substack{\alpha,\beta< -1 \\ \alpha\neq\beta}}\frac{\ketbra{i;\abb\bbb}{i}+\text{h.c.}}{(E_i-E_i^{\abb})(E_i-E_i^{\abb\bbb})}-\sum_{\alpha<-1}
    \frac{\ketbra{i}{i}}{(E_i-E_i^{\abb})^2}+\sum_{\alpha,\beta<-1}\frac{\ketbra{i;\abb}{i;\bbb}}{(E_i-E_i^{\abb})(E_i-E_i^{\bbb})}
    \bigg)\\
    &+J_1h_2\sum_{\substack{\alpha\ge 0 \\ \beta<-1}}\bigg(\frac{(2E_i-E_i^{\ab}-E_i^{\bbb})\ketbra{i;\ab\bbb}{i}+\text{h.c.}}{(E_i-E_i^{\ab})(E_i-E_i^{\bbb})(E_i-E_i^{\ab\bbb})}+\frac{\ketbra{i;\ab}{i;\bbb}+\text{h.c.}}{(E_i-E_i^{\ab})(E_i-E_i^{\bbb})}\bigg)\\
    &+J_1h_2\sum_{\alpha\ge 0}\bigg(
    \bigg(\frac{\bra{i}\Z_0\ket{i}}{(E_i-E_i^{\nop})(E_i-E_i^{\nop\ab})}+
    \frac{\bra{i;\ab}\Z_0\ket{i;\ab}}{(E_i-E_i^{\ab})(E_i-E_i^{\nop\ab})}
    \bigg)(\ketbra{i;\nop\ab}{i}+\text{h.c.})\\
    &+\frac{\bra{i}\Z_0\ket{i}(\ketbra{i;\ab}{i;\nop}+\text{h.c.})}{(E_i-E_i^{\ab})(E_i-E_i^{\nop})}
    \bigg)\\
    &+J_1^2\sum_{\alpha<-1}\bigg(
    \bigg(\frac{(2E_i-E_i^{\nop}-E_i^{\abb})\bra{i}\Z_0\ket{i}}{(E_i-E_i^{\nop})(E_i-E_i^{\abb})(E_i-E_i^{\nop\abb})}
    \bigg)(\ketbra{i;\nop\abb}{i}+\text{h.c.})\\
    &+\frac{\bra{i}\Z_0\ket{i}(\ketbra{i;\abb}{i;\nop}+\text{h.c.})}{(E_i-E_i^{\abb})(E_i-E_i^{\nop})}
    \bigg)
    +\frac{J_1^2}{(E_i-E_i^{\nop})^2}(\ketbra{i;\nop}{i;\nop}-\ketbra{i}{i})
    .
\end{split}
\end{equation}
When we plug the $R_{ii}^{(1)}$s into Eq.~\eqref{ZPT} to get $\Psi^{(1)}$, we find that all of the terms cancel except for those with $\alpha=0$.
This makes sense; this just means that the perturbation needs to flip the edge spin (which was conserved when $\Z_0$ was the SZM) to modify the SZM. Flipping spin zero has an energy cost of $\pm 2J_2$ in the boundary Ising model, so there are no poles at first order.

For $\Psi^{(2)}$, all terms not involving a flip of spin $0$ will cancel.
Furthermore, almost all terms will also cancel given the energy of both excitations together above the unperturbed state is simply the sum  of each separately; specifically, they will cancel if
\begin{equation}
    E_i^{ab}-E_i=(E_i^a-E_i)+(E_i^b-E_i)\label{noninteracting}
\end{equation}
where $a$ and $b$ denote an excitation (i.e.~either flipping one spin on the ferromagnetic side or two adjacent spins on the paramagnetic side).
Here, of course, one of those excitations will be flipping spin 0.
For the boundary Ising model, all pairs of excitations except for adjacent spins flips in the ferromagnetic phase and overlapping pairs of spin flips in the paramagnetic phase obey Eq.~\eqref{noninteracting}, and therefore will not contribute poles where this condition applies.
If we were to flip a sequence of adjacent spins in the ferromagnet starting from spin $0$, however, any domain wall created/destroyed by flipping spin $i$ will be destroyed/created by flipping spin $i+1$, and we can only create or destroy one domain wall on net, changing the energy by $\pm 2J_2$.
Similarly, suppose only spin $i$ is flipped in the paramagnetic phase (the first paramagnetic spin we flip must be spin $-1$ with the boundary term $-J_1\Z_{-1}\Z_0$, as described below).
Then we could flip spin $i$ and $i-1$, evading Eq.~\eqref{noninteracting}, but now only spin $i-1$ is flipped ($i$ having been flipped and flipped back).
Just as with the domain walls, we can only flip one paramagnetic spin on net.
If we were to add some terms to $H_0$ (that didn't change our assumed eigenstates) to make additional pairs of excitations that broke the separability condition \eqref{noninteracting} (for example, next nearest neighbor ZZ interactions) we would get additional poles at $E_i=E_i^{ab}$.

As we implied above, there is an exception to this condition enforcing that terms cancel.
Specifically, it does not apply when one of the excitations involves the boundary term $-J_1\Z_{-1}\Z_0$, since that term measures the sign of the boundary spin $\Z_0$, which flips during any processes which contribute to the perturbation theory.
Instead, the cancellation condition becomes
\begin{equation}
    E_i^{ab}-E_i=(E_i^a-E_i)-(E_i^b-E_i),
\end{equation}
such that the excitations now \emph{subtract} independently to give the total energy above the unperturbed state. In general, this means such terms will not cancel.

For the boundary Ising model, then, we can create one spin flip in the paramagnetic phase and one domain wall in the ferromagnetic phase using $-J_1\Z_{-1}\Z_0$ and $-h_2\X_0$ and then move those excitations further into their respective phases, but we can only change the system's energy by $\pm 2h_1\pm 2J_2$ on net after creating a spin flip/domain wall pair and moving either/both, explaining the lack of additional poles in $\Psi$. 
We could make additional excitations at the boundary after making room by moving away any previous excitations, but after forming $m$ excitations the total change in energy will just be $\pm 2mh_1\pm 2mJ_2$, again generating no new resonances.

We have demonstrated the conditions under which a SZM has a resonance in boundary Ising-like models up to second order in perturbation theory. The key intuition we developed is that a resonance requires not only states of equal energy with the spin of interest flipped, but also a process connecting those two states driven by a sequence of excitations, whose energies depend on each others' presence and the spin of interest. We expect this criterion to be of general importance to SZMs in free and integrable spin chains.

\end{document}